\documentclass[12pt]{article}
\usepackage{geometry}
\usepackage{amsthm, amsmath, amsmath, amssymb, bbm, bm}
\usepackage[round]{natbib}
\usepackage[colorlinks, 
linkcolor=blue,
anchorcolor=blue,
urlcolor=blue,
citecolor=blue]{hyperref}
\usepackage{graphicx}
\usepackage{mathrsfs}
\usepackage{threeparttable}
\usepackage[title]{appendix}
\usepackage{url}
\usepackage{float}
\usepackage{etoolbox}

\usepackage{diagbox}
\usepackage{booktabs}
\usepackage{caption}
\usepackage{subfigure}
\usepackage{listings}
\usepackage{multirow}
\usepackage{rotating}
\usepackage{longtable}
\usepackage{enumitem}
\usepackage{cases}
\usepackage{algorithm}
\usepackage{algorithmic}
\usepackage{ulem}
\usepackage{authblk}
\usepackage{afterpage}
\usepackage{placeins}

\usepackage{xr}
\newtheorem{theorem}{Theorem}[section]

\newtheorem{proposition}{Proposition}[section]

\newtheorem{assumption}{Assumption}

\newtheoremstyle{mycase}{5pt}{5pt}{\upshape}{}{\bfseries}{.}{ }{}\theoremstyle{mycase}

\newtheorem{remark}{Remark}

\newtheoremstyle{myIV}{1.5pt}{1.5pt}{\upshape}{1em}{\bfseries}{.}{ }{}\theoremstyle{myIV}

\newtheoremstyle{mysets}{1.5pt}{1.5pt}{\upshape}{}{\bfseries}{.}{ }{} \theoremstyle{mysets}

\renewcommand{\theequation}{\thesection.\arabic{equation}}
\numberwithin{equation}{section}

\renewcommand{\hat}{\widehat}
\renewcommand{\tilde}{\widetilde}
\renewcommand{\theequation}{\thesection.\arabic{equation}}

\renewcommand{\hat}{\widehat}
\renewcommand{\tilde}{\widetilde}
\newcommand{\Cov}{\mathrm{Cov}}
\newcommand{\Var}{\mathrm{Var}}
\newcommand{\bB}{\boldsymbol{B}}
\newcommand{\bbx}{\boldsymbol{x}}
\newcommand{\bX}{\boldsymbol{X}}
\newcommand{\bbf}{\boldsymbol{f}}

\newcommand{\bbu}{\boldsymbol{u}}

\newcommand{\bbeta}{\bm{\beta}}

\newcommand{\bgamma}{\bm{\gamma}}

\newcommand{\bSigma}{\bm{\Sigma}}

\newcommand{\mbE}{\mathbb{E}}

\newcommand{\tr}{\mathrm{tr}}

\begin{document}
	\allowdisplaybreaks[3] %allow align environment spread page
	
	\title{\bf High-dimensional inference for single-index model with latent factors}

	\author{Yanmei Shi$^{1}$, Meiling Hao$^{2*}$, Yanlin Tang $^{3}$, Heng Lian$^4$, and Xu Guo$^1$\footnote{Corresponding authors: Meiling Hao and Xu Guo; email: meilinghao@uibe.edu.cn and xustat12@bnu.edu.cn}
		\\~
		\\	{\small \it $^{1}$ School of Statistics, Beijing Normal University, Beijing, China}\\
		%	{\small \it $^{2}$ Institute of Statistics and Big Data, Renmin University of China, Beijing, China}\\
		{\small \it $^{2}$ School of Statistics, University of International Business and Economics, Beijing, China}\\
		
		{\small \it $^{3}$ Key Laboratory of Advanced Theory and
			Application in Statistics and Data
			Science–MOE, School of Statistics, East
			China Normal University, Shanghai,
			China}\\
		{\small \it $^{4}$ Department of Mathematics, City University of Hong Kong, Hong Kong, China}\\
	}
	%\author[1]{Yanmei Shi}
	%\author[1]{Xu Guo}
	
	%\affil[1]{\small \it School of Statistics, Beijing Normal University, Beijing, China}
	
	\renewcommand\Authands{, and }

	\date{}
	\maketitle
	
	\vspace{-0.5in}
	
	\begin{abstract}
	Models with latent factors recently attract a lot of attention. However, most investigations focus on linear regression models and thus cannot capture nonlinearity. To address this issue, we propose a novel Factor Augmented Single-Index Model. 	
	We first address the concern whether it is necessary to consider the augmented part by introducing a score-type test statistic. Compared with previous test statistics, our proposed test statistic does not need to estimate the high-dimensional regression coefficients, nor high-dimensional precision matrix, making it simpler in implementation. We also propose a Gaussian multiplier bootstrap to determine the critical value. The validity of our procedure is theoretically established under suitable conditions. We further investigate the penalized estimation of the regression model. With estimated latent factors, we establish the error bounds of the  estimators. Lastly, we introduce debiased estimator and construct confidence interval for individual coefficient based on the asymptotic normality. No moment condition for the error term is imposed for our proposal. Thus our procedures work well when random error follows heavy-tailed distributions or when outliers are present. We demonstrate the finite sample performance of the proposed method through comprehensive numerical studies and its application to an FRED-MD macroeconomics dataset.
\end{abstract}


\section{Introduction}

{The rapid advancement of information technology has brought  significant changes in both data collection and analysis.  Various disciplines, such as economics, social sciences, and genetics, increasingly collect high-dimensional data for comprehensive research and analysis \citep{belloni2012sparse, buhlmann2014high, fan2020statistical}. 	Most existing high-dimensional procedures (for a recent comprehensive review, see for instance \cite{fan2020statistical}) were conducted under the assumption that there are no latent factors associated with both the response variables and covariates. However,  {it is important to acknowledge that this assumption is often violated in real-world.}
	For instance, in genetic studies, the influence of specific DNA segments on gene expression can be confounded by population structure and artifacts in microarray expression \citep{listgarten2010correction}. 
	Similarly, in healthcare research, disentangling the impact of nutrient intake on cancer risk can be confounded by factors such as physical health, social class, and behavioral aspects \citep{fewell2007impact}.
	Failure to account for latent factors in the model {may} lead to biased inferences.
	%, even when using aforementioned debiased methods, potentially yielding incorrect results.
	These examples underscore the importance of considering latent factors within models to ensure accurate and reliable conclusions.
	
	To tackle the issues posed by latent factors, \cite{fanJ2023} proposed the following   Factor Augmented sparse linear Regression Model (FARM):
	\begin{align} 
		Y&=\bbf^{\top}\bgamma+\bbu^{\top}\bbeta+\varepsilon, \label{linear regression models with hidden confounders} \\
		\text {with} \ \ 
		\bbx&=\bB\bbf+\bbu.\label{Factormodel}
	\end{align}
	Here, $Y$  is a response variable, $\bbx$ is a $p$-dimensional covariate vector, $\bbf$ is a $K$-dimensional vector of latent factors, $\bB \in \mathbb{R}^{p \times K}$ is the corresponding factor loading matrix, and $\bbu$ is a $p$-dimensional vector  of idiosyncratic components, which is uncorrelated with $\bbf$. The $\bbeta=(\beta_1,\ldots,\beta_p)^\top\in \mathbb{R}^{p}$ and $\bgamma=(\gamma_1,\ldots,\gamma_K)^\top\in \mathbb{R}^{K}$ are vectors of regression parameters quantifying the contribution of
	$\bbu$  and  $\bbf$, respectively.  The random error $\varepsilon$  satisfies that $\mbE\left(\varepsilon\right)=0$, and is independent of $\bbu$ and $\bbf$.
	In fact, model \eqref{Factormodel} is a commonly used structure for characterizing the  interdependence among features. 
	In this framework, the variables are intercorrelated through a shared set of latent factors.

	Numerous methodologies have been proposed to enable  statistical analysis regarding models with latent factors. \cite{guo2022doubly} introduced a deconfounding approach for conducting statistical inference on individual regression coefficient ${\beta}_{j}, j=1,\ldots,p$, integrating the findings of \cite{cevid2020spectral} with the debiased Lasso method \citep{zhang2014confidence, van2014asymptotically}. \cite{ouyang2023high}
	investigated the inference problem of generalized linear regression models (GLM) with latent factors. \cite{sun2023decorrelating} 
	considered the multiple testing problem for GLM with latent factors. \cite{bing2023inference}  focused on inferring high-dimensional multivariate response regression models with latent factors.
	
	Although the above FARM is powerful to deal with latent factors, it may be not flexible enough to handle the nonlinear relationship between covariates and the response. To capture the nonlinearity, the single-index model (SIM) is usually adopted due to its flexibility
	and interpretability. As a result, the SIM has been the subject of extensive attention and in-depth research in the past decade. Actually, several approaches have been explored for the variable selection problem in high-dimensional SIMs. Examples include  \cite{kong2007variable}, \cite{zhu2009nonconcave}, \cite{wang2012non},  \cite{radchenko2015high}, \cite{plan2016generalized}, and \cite{rejchel2020rank}.  However, limited attention has been paid to discuss SIMs with latent factors.  This motivates us to investigate the high-dimensional single-index
	model with latent factors.
	
	To this end, we consider the following Factor Augmented sparse Single Index Model (FASIM), which integrates both the latent factors and the covariates, 
	\begin{align} \label{FASIM}
		Y&=g(\bbf^{\top}\bgamma+\bbu^{\top}\bbeta,\varepsilon), \notag \\
		\text{with} \ \ \bbx&=\bB\bbf+\bbu.
	\end{align}
	Here the link function $g(\cdot, \cdot)$  is unknown, {and} other variables and parameters in model \eqref{FASIM} remain consistent with {those} defined in model \eqref{linear regression models with hidden confounders}.} When $g(a, b)=a+b$, the above FASIM reduces to the FARM introduced by \cite{fanJ2023}. Since $g(\cdot, \cdot)$ can be unknown, the above FASIM is very flexible and can capture the nonlinearity. 

For the FASIM, the first concern is whether $\bbeta$ is zero or not. Actually if $\bbeta=\boldsymbol{0}$, the model reduces to a single-index factor regression model, which was also considered by \cite{fan2017sufficient},  \cite{jiang2019sufficient}, and \cite{luo2022inverse}. For this important problem, \cite{fanJ2023} considered the maximum of the debiased lasso estimator of $\bbeta$ under FARM. However, their procedure is computationally expensive since it requires to estimate high-dimensional regression coefficients and also high-dimensional precision matrix. In this paper, we first introduce a score-type test statistic, which does not need to estimate high-dimensional regression {coefficients}, nor high-dimensional precision matrix. Hence, our procedure 
is very simple in implementation. We also propose a Gaussian multiplier bootstrap to determine our test statistic’s critical value. The validity of our procedure is theoretically established under suitable conditions. We also give a power analysis for our test procedure. 

When the FASIM is adequate, it is then of importance to estimate the parameters in FASIM. For the SIM, \cite{rejchel2020rank} considered distribution function transformation of the responses, and then estimated the unknown parameters using the Lasso method. Similar procedures have also been investigated by \cite{zhu2009nonconcave} and \cite{wang2012non}. This enhances the model's robustness in scenarios where random errors follow heavy-tailed distributions or when outliers are present. Motivated by their procedures, we also consider the distribution function transformation of the responses and then introduce penalized estimation of unknown parameters. However, it should be emphasized here, different from  \cite{zhu2009nonconcave}, \cite{wang2012non} and \cite{rejchel2020rank}, in our model $\boldsymbol{f}$ and $\boldsymbol{u}$ are unobserved and must be estimated firstly. This would give additional technical difficulty. In this paper, we establish the estimation error bounds of our introduced penalized estimators under mild assumptions. Notably, no moment condition is required for the error term in the FASIM.

Lastly, we investigate the construction of confidence interval for each regression coefficient in the FASIM. Due to the inherent bias, the penalized estimator cannot be directly used in statistical inference. 
\cite{eftekhari2021inference} investigated the inference problem of the SIM by adopting the debiasing technique \citep{zhang2014confidence, van2014asymptotically}. However, their procedure cannot handle latent factors. To this end, we introduce debiased estimator for the FASIM and also establish its corresponding asymptotic normality. Compared with \cite{eftekhari2021inference}, our procedure does not need sample-splitting and is robust to outliers.

The remainder of this paper is structured as follows. In Section \ref{model construction and factor estimate}, we delve into the reformulation of the FASIM and the estimation of the latent factors. In Section \ref{Adequacy test of factor model}, we develop a powerful test procedure for testing whether $\bbeta=\boldsymbol{0}$. In Section \ref{Estimation and debiased Lasso inference}, we consider the regularization estimation of $\bbeta$ and establish the $\ell_{1}$ and $\ell_{2}$-estimation error bounds for this estimation. Further we introduce debiased estimator and construct confidence interval for each coefficient. We present the findings of our simulation studies in Section \ref{Numerical studies section} and provide an analysis of real data in Section \ref{Real data analysis section} to assess the performance and effectiveness of the proposed approach. Conclusions and discussions are presented in Section \ref{Conclusions and discussions section}. Proofs of the main theorems are provided in the Appendix. Proofs of related technical Lemmas are attached in the  Supplementary Material.

\textbf{Notation.} Let $\mathbb{I}(\cdot)$  denote the indicator function. For a vector $\boldsymbol{a}=\left(a_{1}, \ldots, a_{m}\right)^{\top}\in \mathbb{R}^{m}$, we denote its $\ell_{q}$ norm as $\|\boldsymbol{a}\|_{q}=\left(\sum_{\ell=1}^{m}|a_{\ell}|^{q}\right)^{1/q}, \ 1\leq q< \infty$, $\|\boldsymbol{a}\|_{\infty}=\max_{1\leq \ell \leq m}|a_{\ell}|$,  and $\|\boldsymbol{a}\|_{0}=\sum_{\ell=1}^{m}\mathbb{I}(a_{\ell}\neq 0)$.
For any integer $m$, we define $[m]=\{1,\ldots, m\}$.
The Orlicz norm  of a scalar random variable $X$ is defined as $\|X\|_{\psi_{2}}=\inf{\{c>0: \mathbb{E}\exp(X^{2}/c^{2})\leq 2\}}$.
For a random vector $\boldsymbol{x}\in \mathbb{R}^{m}$, we define its  Orlicz norm  as $\|\boldsymbol{x}\|_{\psi_{2}}=\sup_{\|\boldsymbol{c}\|_{2}=1}\|\boldsymbol{c}^{\top}\boldsymbol{x}\|_{\psi_{2}}$.
Furthermore, we use $\boldsymbol{I}_{K}$,  $\boldsymbol{1}_{K}$ and $\boldsymbol{0}_{K}$ to denote the identity matrix in $\mathbb{R}^{K \times K}$, a vector of dimensional $K$ with all elements being $1$ and all elements being $0$, respectively.
For a matrix $\boldsymbol{A}={(A_{jk})}$, 
{ $\boldsymbol{a}_{i}$ represents {the} $i$-th row of $\boldsymbol{A}$, and $\boldsymbol{A}_{j}$ represents {the} $j$-th column of $\boldsymbol{A}$.}
We define $\|\boldsymbol{A}\|_{\mathbb{F}}=\sqrt{\sum_{jk}{A}_{jk}^{2}}$, $\|\boldsymbol{A}\|_{\max}=\max_{j,k}|A_{jk}|$, $\|\boldsymbol{A}\|_{\infty}=\max_{j}\sum_{k}|A_{jk}|$, $\|\boldsymbol{A}\|_{1}=\max_{k}\sum_{j}|A_{jk}|$ and $\|\boldsymbol{A}\|_{\text{sum}}=\sum_{j,k}|A_{jk}|$
to be its Frobenius norm, element-wise max-norm, matrix $\ell_{\infty}$-norm, matrix $\ell_{1}$-norm and  the element-wise sum-norm, respectively.
Besides, we use $\lambda_{\min}(\boldsymbol{A})$ and $\lambda_{\max}(\boldsymbol{A})$  to denote the minimal and maximal eigenvalues of $\boldsymbol{A}$, respectively.
We use $|\mathcal{A}|$ to denote the cardinality of a set  $\mathcal{A}$.
For two positive sequences $\{a_{n}\}_{n \geq 1}$, $\{b_{n}\}_{n \geq 1}$, we write $a_{n}=O(b_{n})$ if there exists a positive constant $C$ such that $a_{n}\leq C\cdot b_{n}$, and we write $a_{n}=o(b_{n})$ if $a_{n}/b_{n}\rightarrow 0$. 
Furthermore, if $a_{n}=O(b_{n})$ is satisfied, we write $a_n\lesssim b_n$. If
$a_n\lesssim b_n$ and $b_n\lesssim a_n$, we  write it as $a_n\asymp b_n$ for short.
In addition, $a_{n}=O_{\mathbb{P}}(b_{n})$ and $a_{n}=o_{\mathbb{P}}(b_{n})$ have similar meanings as above except that the relationship of $a_{n}/b_{n}$ holds with high probability.  The parameters $c, c_{0},  C, C_{1}, C_{2}$ and $K'$  appearing in
%this paper differ from each other by at most an absolute constant factor.
this paper are all positive constants.

\section{Factor augmented  single-index model} \label{model construction and factor estimate}
In this section, we investigate  the reformulation of the FASIM and the estimation of the latent factors, as well as their properties.

\subsection{Reformulation of FASIM}
The unknown smooth function $g(\cdot, \cdot)$ makes the estimation of the FASIM challenging. To address this concern, we reformulate the FASIM as a linear regression model with transfomed response. With this transformation, we estimate the parameters avoiding estimating the unknown function $g(\cdot, \cdot)$. 

Without loss of generality, we assume   that $\mbE(\boldsymbol{x})=\boldsymbol{0}$, and $\boldsymbol{\Sigma}=\Cov({\boldsymbol{x}}) $ is a positive definite matrix. Let $\boldsymbol{\sigma}_{\boldsymbol{u}h}=\Cov\{\boldsymbol{u}, h(Y)\}$ and $\boldsymbol{\sigma}_{\boldsymbol{f}h}=\Cov\{\boldsymbol{f}, h(Y)\}$ for a given transformation function $h(\cdot)$ of the response. Define  {$\boldsymbol{\beta}_{h}=\mathbb{E}\left(\boldsymbol{u}\boldsymbol{u}^{\top}\right)^{-1}\boldsymbol{\sigma}_{{\boldsymbol{u}}h}$}, $\boldsymbol{\gamma}_{h}=\mathbb{E}\left(\boldsymbol{f}\boldsymbol{f}^{\top}\right)^{-1}\boldsymbol{\sigma}_{\boldsymbol{f}h}$, $\boldsymbol{\eta}_{h}=(\boldsymbol{\beta}_{h}^{\top}, \boldsymbol{\gamma}_{h}^{\top})^{\top}$, $\boldsymbol{\eta}=(\boldsymbol{\beta}^{\top}, \boldsymbol{\gamma}^{\top})^{\top}$ and $\boldsymbol{v}=(\boldsymbol{u}^{\top}, \boldsymbol{f}^{\top})^{\top}$. 
In the context of the SIM framework, the following linear expectation condition is commonly assumed.

\begin{proposition}\label{P1}
	Assume that $\mathbb{E}\left(\boldsymbol{v} \mid \boldsymbol{v}^{\top}\boldsymbol{\eta}\right)$ is a linear function of $\boldsymbol{v}^{\top}\boldsymbol{\eta}$. Then $\boldsymbol{\eta}_{h}$ is proportional to $\boldsymbol{\eta}$, that is $\boldsymbol{\eta}_{h}=\kappa_{h} \times \boldsymbol{\eta}$ for some constant $\kappa_{h}$.
\end{proposition}

Proposition \ref{P1} is from Theorem 2.1 of \cite{li1989regression}. The condition in this proposition is referred to as the linearity condition (LC) for predictors. It is
satisfied when $\boldsymbol{v}$  follows an elliptical distribution, a commonly used assumption in the sufficient dimension reduction literature  \citep{li1991sliced, cook2005sufficient}. \cite{hall1993almost} showed that  the LC holds approximately to many high-dimensional settings. Throughout  the paper, we assume  $\kappa_{h}\neq 0$, which is a relatively mild assumption. In fact, when $h(\cdot)$  is monotone  and $g(\cdot, \cdot)$ is monotonic with respect to the first argument, 
{this} assumption is anatomically satisfied.

Following from the definition of $\boldsymbol{\beta}_{h}$ and $\boldsymbol{\gamma}_{h}$,
our model can be rewritten as:
\begin{align} \label{equivalent form}
	h(Y)&=\boldsymbol{x}^{\top}\boldsymbol{\beta}_{h}+\boldsymbol{f}^{\top}\boldsymbol{\varphi}_{h}+{e}_{h},
\end{align}
where $\boldsymbol{\varphi}_{h}=\boldsymbol{\gamma}_{h}-\boldsymbol{B}^{\top}\boldsymbol{\beta}_{h}$, {and the error term $e_{h}$  satisfies $\mathbb{E}(e_{h})={{0}}$},  $\mathbb{E}(e_{h}\boldsymbol{u})={\boldsymbol{0}_{p}}$ and
$\mathbb{E}(e_{h}\boldsymbol{f})={\boldsymbol{0}_{K}}$. This implies that we now recast the FASIM as a FARM with transformed response. Actually for identification of the FASIM, it is usually assumed that $\|\boldsymbol{\eta}\|_2=1$ and its first element is positive. Thus the estimation of the direction of $\boldsymbol{\eta}$ is sufficient for the FASIM. The proportionality between $\boldsymbol{\eta}$ and $\boldsymbol{\eta}_h$, along with the previous mentioned transformed linear regression model (\ref{equivalent form}), reduces the difficulty of analysis. 

In practice, we need to choose an appropriate transformation function $h(\cdot)$. It's worth mentioning that the  procedure  in \cite{eftekhari2021inference} essentially works with $h(Y)=Y$. Motivated by \cite{zhu2009nonconcave} and \cite{rejchel2020rank}, we set $h(Y)=F(Y)-1/2$, where $F(Y)$ is the distribution function of $Y$. This specific choice could make our procedures be robust 
against outliers and heavy tails.  Actually with equation (\ref{equivalent form}), given the widely imposed sub-Gaussian assumption on the predictors, the boundedness of $F(Y)-1/2$ would lead the transformed error term $e_h$
being sub-Gaussian, even if the original error term $\varepsilon$
comes from Cauchy distribution. The response-distribution transformation is preferred due to some additional reasons which will be discussed later.

\subsection{Factor estimation}
Throughout the paper, we assume that the data $\{\boldsymbol{x}_{i}, \boldsymbol{f}_{i}, Y_{i},e_{hi}\}_{i=1}^{n}$ are independent and identically distributed (i.i.d.) copies of $\{\boldsymbol{x}, \boldsymbol{f}, Y,e_{h}\}$. {Let  $\boldsymbol{X}=(\boldsymbol{x}_{1}, \ldots, \boldsymbol{x}_{n})^{\top}$, $\boldsymbol{F}=(\boldsymbol{f}_{1}, \ldots, \boldsymbol{f}_{n})^{\top}$, $\boldsymbol{Y}=(Y_{1}, \ldots, Y_{n})^{\top}$ and $\boldsymbol{e}_{h}=({e}_{h1},\ldots,{e}_{hn})^{\top}$. We consider the high-dimensional scenario where the dimension $p$ of the observed covariate vector $\boldsymbol{x}$  can be much larger than the sample size $n$. 
	For the FASIM, we first need to estimate the latent factor vector $\boldsymbol{f}$ since only the predictor vector ${\boldsymbol{x}}$ and the response $Y$ are observable. To address this issue, we impose an identifiability assumption similar to that in \cite{FanJ2013}. That is  
	\begin{align}
		\Cov(\boldsymbol{f})=\boldsymbol{I}_{K}, \ \ \text{and} \ \ \boldsymbol{B}^{\top}\boldsymbol{B} \ \ \text{is  \  diagonal}. \notag
	\end{align}
	Consequently, the constrained least squares estimator of $(\boldsymbol{F},\boldsymbol{B})$ based on ${\bX}$ is given as
	\begin{align}
		(\hat{\boldsymbol{F}},\hat{\boldsymbol{B}})=&\arg\min\limits_{\boldsymbol{F}\in \mathbb{R}^{n\times K},~\boldsymbol{B}\in \mathbb{R}^{p\times K}}\|{\bX}-\boldsymbol{F}\boldsymbol{B}^{\top}\|_{\mathbb{F}}^{2}, \\ \notag
		\text{subject} \ \  \text{to}& \  \frac{1}{n}{\boldsymbol{F}}^{\top}\boldsymbol{F}=\boldsymbol{I}_{K} \ \ \text{and} \ \ {\boldsymbol{B}}^{\top}\boldsymbol{B} \ \ \text{is} \ \  \text{diagonal}. \notag
	\end{align}
	Let $\boldsymbol{U}=(\boldsymbol{u}_{1}, \ldots, \boldsymbol{u}_{n})^{\top}$, {with $\boldsymbol{u}_{i}=\boldsymbol{x}_{i}-\boldsymbol{B}\boldsymbol{f}_{i}, i=1,\ldots,n$}. {Denote $\boldsymbol{\Sigma}_{\boldsymbol{u}}=\Cov(\boldsymbol{u})$.} 
	
	Elementary manipulation yields that the columns of $\hat{\boldsymbol{F}}/\sqrt{n}$ are the eigenvectors corresponding to the largest $K$ eigenvalues of the matrix ${\bX}{{\bX}}^{\top}$ and $\hat{\boldsymbol{B}}={\bX}^{\top}{\hat{\boldsymbol{F}}}({\hat{\boldsymbol{F}}}^{\top}\hat{\boldsymbol{F}})^{-1}=n^{-1}{\bX}^{\top}{\hat{\boldsymbol{F}}}$.
	Then the estimator of $\boldsymbol{U}$ is  $$\hat{\boldsymbol{U}}={\bX}-\hat{\boldsymbol{F}}{\hat{\boldsymbol{B}}}^{\top}=\left(\boldsymbol{I}_{n}-\frac{1}{n}\hat{\boldsymbol{F}}{\hat{\boldsymbol{F}}}^{\top}\right){\bX},$$
	see \cite{FanJ2013}.  Since $K$  is related to the number of spiked eigenvalues of ${\bX}{\bX}^{\top}$, it is usually small. Therefore, we treat $K$ as a fixed constant as suggested by \cite{fanJ2023}.  {{Additionally}, let $\hat{\boldsymbol{\Sigma}}$, $\hat{\boldsymbol{\Lambda}}=\text{diag}\left(\hat{\lambda}_{1},\ldots, \hat{\lambda}_{K}\right)$ and $\hat{\boldsymbol{\Gamma}}=\left(\hat{\boldsymbol{\zeta}}_{1},\ldots, \hat{\boldsymbol{\zeta}}_{K}\right)$ be suitable estimators of the covariance matrix $\boldsymbol{\Sigma}$ of $\boldsymbol{x}$, the matrix consisting of its leading $K$ eigenvalues $\boldsymbol{\Lambda}=\text{diag}(\lambda_{1},\ldots, \lambda_{K})$ and the matrix consisting of their corresponding $K$ orthonormalized eigenvectors $\boldsymbol{\Gamma}=(\boldsymbol{\zeta}_{1},\ldots, \boldsymbol{\zeta}_{K})$, respectively.} 
	
	We proceed by presenting the regularity assumptions imposed in seminal works about factor analysis, such as \cite{Bai2003} and \cite{FanJ2013,fanJ2023}.
	\begin{assumption} \label{factorassumption} 
		We make the following assumptions. 		
		\begin{itemize}
			
			\item[(i)] There exists a positive constant $c_{0}$ such that $\|\boldsymbol{f}\|_{\psi_{2}}\leq c_{0}$ and $\|\boldsymbol{u}\|_{\psi_{2}}\leq c_{0}$. In addition, $\mathbb{E}(U_{ij})=\mathbb{E}(F_{ik})=0, \ i=1,\ldots,n, \ j=1,\ldots,p, \  k=1,\ldots,K$.
			\item[(ii)] There exists a constant $\tau >1$ such that $p/\tau\leq \lambda_{min}(\boldsymbol{B}^{\top}\boldsymbol{B})\leq \lambda_{\max}(\boldsymbol{B}^{\top}\boldsymbol{B})\leq p\tau$. 
			\item[(iii)]  
			There exists a constant $\Upsilon>0$ such that $\|\boldsymbol{B}\|_{\max}\leq \Upsilon$ and $$\mathbb{E}|\boldsymbol{u}^{\top}\boldsymbol{u}-\tr(\boldsymbol{\Sigma}_{\boldsymbol{u}})|^{4}\leq \Upsilon p^{2}.$$ 
			\item[(iv)] There exists a positive constant $\kappa<1$ such that $\kappa \leq \lambda_{\min}(\boldsymbol{\Sigma}_{\boldsymbol{u}}), \ \lambda_{\max}(\boldsymbol{\Sigma}_{\boldsymbol{u}})$, $\|\boldsymbol{\Sigma}_{\boldsymbol{u}}\|_{1}\leq 1/\kappa$, and $\min_{1\leq k,l\leq p} \Var({U}_{ik}{U}_{il})\geq \kappa$.
			
		\end{itemize}
	\end{assumption}
	\begin{assumption} (Initial pilot estimators). \label{Loadings and initial pilot estimators}
		Assume that 
		$\hat{\boldsymbol{\Sigma}}$, $\hat{\boldsymbol{\Lambda}}$ and $\hat{\boldsymbol{\Gamma}}$ satisfy $\|\hat{\boldsymbol{\Sigma}}-\boldsymbol{\Sigma}\|_{\max}=O_{\mathbb{P}}\{\sqrt{(\log p)/n}\}$, $\|(\hat{\boldsymbol{\Lambda}}-\boldsymbol{\Lambda})\hat{\boldsymbol{\Lambda}}^{-1}\|_{\max}=O_{\mathbb{P}}\{\sqrt{(\log p)/n}\}$, and $\|\hat{\boldsymbol{\Gamma}}-\boldsymbol{\Gamma}\|_{\max}=O_{\mathbb{P}}\{\sqrt{(\log p)/(np)}\}$.
	\end{assumption}
	\begin{remark}
		Assumption \ref{Loadings and initial pilot estimators} is taken from \cite{bayle2022factor}. This assumption holds in various scenarios of interest, such as for the sample covariance matrix under sub-Gaussian distributions \citep{FanJ2013}. Moreover,  the estimators like the marginal and spatial Kendall’s tau \citep{fan2018large}, and the elementwise adaptive Huber estimator \citep{fan2019robust} satisfy this assumption.
	\end{remark}
	
	We summarize the theoretical results related to consistent factor estimation in 
	Lemma 1 in Supplementary Material, which directly follows from Proposition 2.1 in \cite{fanJ2023} and Lemma 3.1 in \cite{bayle2022factor}.
	
	In practice, the number of latent factors {$K$} is often unknown, and determining $K$ in a data-driven way is a crucial challenge. Numerous methods have been introduced in the literature to estimate the value of $K$ \citep{BaiandNg2002, Lam2012, Ahn2013, Fan2022}.
	In this paper, we employ the ratio method for our numerical studies \citep{Luo2009,Lam2012, Ahn2013}. Let $\lambda_{k}({\bX}{\bX}^{\top})$ be the $k$-th largest eigenvalue of ${\bX}{\bX}^{\top}$.
	The number of factors can
	be consistently estimated  by
	\begin{align}
		\hat{K}=\arg\max\limits_{k\leq K_{\max}}\frac{\lambda_{k}({\bX}{\bX}^{\top})}{\lambda_{k+1}({\bX}{\bX}^{\top})}, \notag
	\end{align}
	where $1 \leq K_{\max} \leq n$ is a prescribed upper bound for $K$. In  our subsequent theoretical analysis, we treat $K$  as known. All the theoretical results remain
	valid conditioning on that  $\hat{K}$ is a consistent estimator  of $K$. 
	
	\section{Adequacy test of factor model} \label{Adequacy test of factor model}
	In this section, we aim to assess the adequacy of the factor model and determining whether FASIM \eqref{FASIM} can serve as an alternative  to $Y=g(\boldsymbol{f}^{\top}\boldsymbol{\gamma},\varepsilon)$. The primary question of interest pertains to the following hypothesis: 
	\begin{align} \label{hyporhesis test problem}
		H_{0}: \boldsymbol{\beta}_h=\boldsymbol{0} \ \ \text{versus} \ \ H_{1}: \boldsymbol{\beta}_h\neq \boldsymbol{0}.
	\end{align}
	From Proposition \ref{P1}, the null hypothesis is also equivalent to $H_{0}: \boldsymbol{\beta}=\boldsymbol{0}$. 
	
	\subsection{ Factor-adjusted score-type test}
	In this subsection, we develop a factor-adjusted score type test (FAST) and derive its Gaussian approximation result. For the adequacy testing problem, \cite{fanJ2023} considered the maximum of the debiased lasso estimator of $\bbeta$ under FARM. However, their procedure requires to estimate high-dimensional regression coefficients and also high-dimensional precision matrix, and thus is computationally expensive. While our proposed FAST does not need to estimate high-dimensional regression {coefficients}, nor high-dimensional precision matrix. Hence, Our procedure is straightforward to implement, saving both computation time and computational resources.
	
	Under the null hypothesis in (\ref{hyporhesis test problem}), we have $$\mbE\left[\left\{F(Y)-{\frac{1}{2}}-\boldsymbol{f}^\top\bgamma_h\right\}\boldsymbol{u}\right]
	=\mbE[e_h\boldsymbol{u}]=\boldsymbol{0}.$$
	While under alternative hypothesis $H_1$, we have
	$$\mbE\left[\left\{F(Y)-{\frac{1}{2}}-\boldsymbol{f}^\top\bgamma_h\right\}\boldsymbol{u}\right]
	=\mbE\{\boldsymbol{u}(e_h+\boldsymbol{u}^\top\bbeta_h)\}=\bSigma_{\boldsymbol{u}}\bbeta_h\neq \boldsymbol{0}.$$
	This observation motivates us to consider the following individual test statistic 
	\begin{align}\label{Tn definition}
		{T}_{nj}=\frac{1}{\sqrt{n}}\sum\limits_{i=1}^{n}\left[\left\{F_{n}(Y_{i})-\frac{1}{2}-\hat{\boldsymbol{f}}_{i}^{\top}\hat{\boldsymbol{\gamma}}_{h}\right\}\hat{U}_{ij}\right].
	\end{align}
	Here we  utilize  the empirical distribution $F_{n}(y)=n^{-1}\sum_{i=1}^{n}\mathbb{I}(Y_{i}\leq y)$ as an estimator of $F(Y)$.  In the empirical distribution function, the term 
	$\sum_{j=1}^n \mathbb{I}(Y_j\leq Y_i)$ is the rank of $Y_i$. Since statistics with ranks such as Wilcoxon test and the Kruskall-Wallis ANOVA test, are well-known to be robust, this intuitively explains why our procedures with response-distribution transformation function would be robust with respect to outliers in response. We consider the least squares estimator $\hat{\boldsymbol{\gamma}}_{h}=(\hat{\boldsymbol{F}}^{\top}\hat{\boldsymbol{F}})^{-1}\hat{\boldsymbol{F}}^{\top}\left\{F_{n}(\boldsymbol{Y})-{1}/{2}\right\}$. Denote ${\boldsymbol{T}}_{n}=({T}_{n1},\ldots,{T}_{np})^{\top}$. To test the null hypothesis $H_{0}: \boldsymbol{\beta}_{h}=\boldsymbol{0}$, we consider $\ell_{\infty}$ norm of ${\boldsymbol{T}}_{n}$. 
	That is,
	\begin{align}\label{max type standardized statistic}
		M_{n}=\|{\boldsymbol{T}}_{n}\|_{\infty}.
	\end{align}
	It is clear that in the FAST statistic $M_n$ defined above, we only need to estimate a low-dimensional parameter $\boldsymbol{\gamma}_h$, and no high-dimensional parameters are required. 
	{In addition, since FAST does not rely on the estimation of the precision matrix, we can avoid assumptions about the $\ell_{\infty}$ norm of the precision matrix, as demonstrated in \cite{fanJ2023}.}
	Further define $S_{nj}$ as follows:
	\begin{align} \label{decorrelated score function}
		S_{nj}=\frac{1}{\sqrt{n}}\sum\limits_{i=1}^{n}\left[\left\{F(Y_{i})-\frac{1}{2}-\boldsymbol{f}_{i}^{\top}\boldsymbol{\gamma}_{h}\right\}U_{ij}+m_{j}(Y_{i})\right],
	\end{align}
	where  $m_{j}(y)=\mathbb{E}[(X_{1j}-\boldsymbol{f}_{1}^{\top}\boldsymbol{b}_{j})\{\mathbb{I}(Y\geq y)-F(Y)\}]$.
	Denote  %$\boldsymbol{m}({Y})=\{m_{1}({Y}),\ldots,m_{p}({Y})\}^{\top}$,  
	${\boldsymbol{S}}_{n}=({S}_{n1},\ldots,{S}_{np})^{\top}$ and $\boldsymbol{\Omega}^{*}=\Cov({\boldsymbol{S}}_{n})$. {Let $\sigma_{j}^{2}=\boldsymbol{\Omega}_{jj}^{*}, \ j=1,\ldots,p$.} It could be shown that ${T}_{nj}={S}_{nj}+o_{\mathbb{P}}(1)$. 
	\begin{assumption} \label{Assumption score}
		\item[(i)] $\{\log (np)\}^{7}/n=o(1)$.
		\item[(ii)] {The $\min_{1 \leq j \leq p} \sigma_{j}^{2}$ is bounded away from zero.}
	\end{assumption}
	Assumption \ref{Assumption score} is mild and frequently employed in high-dimensional settings, 
	which is the technical requirement to bound the difference between $M_{n}$ and $\|\boldsymbol{\mathrm{Z}}\|_{\infty}$, where $\boldsymbol{\mathrm{Z}} \sim \mathrm{N}_{p}(\boldsymbol{0}, \boldsymbol{\Omega}^*)$. Specifically, condition (i) imposes suitable restriction on the growth rate of $p$ and is commonly used in the high-dimensional inference literature,  see \cite{zhang2017simultaneous} and \cite{dezeure2017high}. Condition (ii) 
	imposes a boundedness restriction on the second moment of $S_{nj}$, which is also assumed in  \cite{ning2017general}.
	
	\begin{theorem} \label{Score Gaussianapproximationtheoremresult}
		Suppose that %$\|\boldsymbol{\gamma}_{h}\|_{\infty}$ is bounded,   
		Assumptions \ref{factorassumption}-\ref{Assumption score} and LC hold. Under the null hypothesis, that is, $\boldsymbol{\beta}_{h}=\boldsymbol{0}$, we have 
		\begin{align}
			\lim\limits_{n  \rightarrow \infty} \sup \limits_{t \in \mathbb{R}}\left|\Pr(M_{n}\leq t )-\Pr\left(\left\|\boldsymbol{\mathrm{Z}}\right\|_{\infty} \leq t\right)\right|=0,
		\end{align}
		where $\boldsymbol{\mathrm{Z}}\sim \mathrm{N}_{p}\left(\boldsymbol{0}, \boldsymbol{\Omega}^{*}\right)$.
	\end{theorem}
	
	Theorem \ref{Score Gaussianapproximationtheoremresult} indicates that our test statistic can be approximated by the maximum of a high-dimensional Gaussian vector under mild conditions. Based on this,  we can reject the null hypothesis $H_{0}$ at the significant level $\alpha$ if and only if $M_{n} > c_{1-\alpha}$, where $c_{1-\alpha}$ is the $(1-\alpha)$-th quantile of the distribution of $\|\boldsymbol{\mathrm{Z}}\|_{\infty}$.
	
	As demonstrated by many authors, such as \cite{tony2014two} and \cite{ma2021global}, the distribution of $\|\boldsymbol{\mathrm{Z}}\|_{\infty}$ can be asymptotically approximated by the Gumbel distribution. However, this generally requires restrictions on the covariance matrix of $\boldsymbol{\mathrm{Z}}$. For instance, it {requires} that the eigenvalues of $\boldsymbol{\Omega}^{*}$ are uniformly bounded. In addition, the critical value obtained from the Gumbel distribution may not work well in practice since this weak convergence is typically slow \citep{zhang2017simultaneous}. Instead of adopting Gumbel distribution, we consider using bootstrap to approximate the distribution of $M_{n}$. 
	
	\subsection{Gaussian multiplier bootstrap}
	Given that  
	$\boldsymbol{\Omega}^{*}$ is unknown, we employ the Gaussian multiplier bootstrap to derive the critical value  $c_{1-\alpha}$. The procedures and theoretical properties of the Gaussian multiplier bootstrap are outlined in this subsection.
	\begin{itemize}
		\item [1.] Generate  i.i.d.  random variables $\mathrm{N}_{1},\ldots, \mathrm{N}_{n} \sim \mathrm{N}(0,1)$ independent of the observed dataset $\mathcal{D}=\{\boldsymbol{x}_{1},\ldots,\boldsymbol{x}_{n}, Y_{1},\ldots,Y_{n}\}$,    and compute
		\begin{align} \label{bootstrap statistic score}
			\hat{\mathcal{G}}=\max\limits_{j \in [p]}\left|\frac{1}{\sqrt{n}}\sum\limits_{i=1}^{n}\left[\left\{F_{n}(Y_{i})-\frac{1}{2}-\hat{\boldsymbol{f}}_{i}^{\top}\hat{\boldsymbol{\gamma}}_{h}\right\}\hat{U}_{ij}+\hat{m}_{j}(Y_{i})\right]\mathrm{N}_{i}\right|.
		\end{align}
		Here $\hat{m}_{j}(y)={n}^{-1}\sum_{i=1}^{n}\left(X_{ij}-\hat{\boldsymbol{f}}_{i}^{\top}\hat{\boldsymbol{b}}_{j}\right)\{\mathbb{I}(Y_{i}\geq y)-F_{n}(Y_{i})\}$. %Denote  $\hat{\boldsymbol{m}}({Y})=\{\hat{m}_{1}({Y}),\ldots,\hat{m}_{p}({Y})\}^{\top}$.
		\item[2.]Repeat the first step independently for $B$ times to obtain $\hat{\mathcal{G}}_{1},\ldots,\hat{\mathcal{G}}_{B}$. 
		Approximate the critical value $c_{1-\alpha}$ via the $(1-\alpha)$-th quantile of the empirical distribution of the bootstrap statistics:
		\begin{align} \label{critical value estimate score}
			\hat{c}_{1-\alpha}=\inf \left\{t \geq0: \frac{1}{B}\sum\limits_{b=1}^{B}\mathbb{I}(\hat{\mathcal{G}}_{b} \leq t) \geq 1-\alpha\right\}.  
		\end{align} 
		\item[3.]
		We reject the null hypothesis $H_{0}$ if and only if  
		\begin{align} \label{test process score}
			M_{n} \geq \hat{c}_{1-\alpha}.
		\end{align}
	\end{itemize}
	\begin{theorem} \label{the validity of the  bootstrap procedure score}
		Suppose that conditions in Theorem \ref{Score Gaussianapproximationtheoremresult} are satisfied.
		%and Assumption \ref{consistency property assumption} holds, 
		Under the null hypothesis, we have 
		\begin{align}
			\sup\limits_{x>0}\left|\Pr\left(M_{n}\leq x\right)-\Pr(\hat{\mathcal{G}} \leq x|\mathcal{D})\right|\rightarrow 0. \notag 
		\end{align}
		%where $\Pr^{*}(\cdot)=\Pr(\cdot|\mathcal{D})$ denotes the conditional probability.  
	\end{theorem}
	
	Theorem \ref{the validity of the  bootstrap procedure score} demonstrates the validity of the proposed
	bootstrap procedure, grounded in the Gaussian approximation theory established in Theorem \ref{Score Gaussianapproximationtheoremresult}. Based on this result, we can define our decision rule as follows:
	\begin{align*} 
		\psi_{\infty, \alpha}=\mathbb{I}\left(M_{n} >  \hat{c}_{1-\alpha}\right). 
	\end{align*} 
	The null hypothesis $H_{0}$ is rejected if and only if $\psi_{\infty, \alpha}=1$.
	
	\subsection{Power analysis}
	We next  consider the asymptotic power analysis of the  $M_{n}$.  { To demonstrate the efficiency of the test statistic, we consider the following local alternative parameter family for $ \boldsymbol{\beta}_{h}$ under $H_1$.
		\begin{align}
			\boldsymbol{\Delta}(c)=\left\{\boldsymbol{\beta}_{h} \in \mathbb{R}^{p}: \max\limits_{j \in [p]}\left|{\beta_{hj}}\right| \geq \sqrt{c\frac{\log p}{n}}\right\},
		\end{align}
		where $c$ is a positive constant. The  $s=\|\boldsymbol{\beta}_{h}\|_{0}$ quantifies the sparsity of the parameter $\boldsymbol{\beta}_{h}$.} Denote 
	$\mu_j=\mathbb{E}(U_{ij}^2)$.
	\begin{assumption} \label{assumption power}
		Suppose that  $\boldsymbol{\Sigma}_{\boldsymbol{u}}=\Cov(\boldsymbol{u})$ is a diagonal matrix and $\mu_{j}$ is bounded away from zero.
	\end{assumption}
	Assumption \ref{assumption power} outlines the diagonal structure of the covariance matrix  $\boldsymbol{\Sigma}_{\boldsymbol{u}}$, a common assumption in factor analysis \citep{kim1978factor}. We should also note that this diagonality of $\boldsymbol{\Sigma}_{\boldsymbol{u}}$ is only imposed here for simplify the power analysis.

	\begin{theorem} \label{power theorem}
		Suppose that %$\|\boldsymbol{\gamma}_{h}\|_{\infty}$ is bounded,   
		Assumptions \ref{factorassumption}-\ref{assumption power} and LC    hold.   %\hao{If so, why you introduce assumption 4 here???}
		%Under the conditions of Theorem \ref{}, Assumption \ref{factorassumption}-\ref{uncorerlated assumption of u} \hao{still need 3?????},  
		For the test  statistic $M_{n}$ defined in \eqref{max type standardized statistic}, if $s=o({n}/\log p)$,  we have 
		\begin{align}
			\lim\limits_{(n,p) \rightarrow \infty} \inf \limits_{\boldsymbol{\beta}_{h} \in \boldsymbol{\Delta}(2+\varrho_{0})}\Pr\left(M_{n} \geq \hat{c}_{1-\alpha}\right)=1.
		\end{align}
		where  $\varrho_{0}$ is a positive constant.
	\end{theorem}
	
	Theorem \ref{power theorem} suggests that our test procedure maintains high power even when only a few components of $\boldsymbol{\beta}_{h}$ have magnitudes larger than $\sqrt{(2+\varrho_{0})(\log p)/{n}}$. Thus, our testing procedure is powerful against sparse alternatives. Notably, this separation rate represents the minimax optimal rate for local alternative hypotheses, as discussed in \cite{verzelen2012minimax}, \cite{tony2014two}, \cite{zhang2017simultaneous}, and \cite{ma2021global}.
	
	\section{Estimation and inference of FASIM} \label{Estimation and debiased Lasso inference}
	When the null hypothesis is rejected, indicating that the factor model is inadequate, we need to consider FASIM. In this section, we aim to investigate the estimation and inference of the FASIM.  
	
	\subsection{Regularization estimation}
	In the high-dimensional regime,
	we employ $\ell_1$ penalty \citep{tibshirani1996regression} to estimate
	the unknown parameter vectors $\boldsymbol{\beta}_{h}$
	and $\boldsymbol{\gamma}_{h}$ in model \eqref{equivalent form}:
	\begin{align} \label{empiricalestimator}
		(\hat{\boldsymbol{\beta}}_{h},\hat{\boldsymbol{\gamma}}_{h})=\arg\min\limits_{\boldsymbol{\beta}\in \mathbb{R}^{p},\boldsymbol{\gamma}\in \mathbb{R}^{K}} \left[\frac{1}{2n}\sum\limits_{i=1}^{n}\left\{F_n(Y_{i})-\frac{1}{2}-\hat{\boldsymbol{u}}_{i}^{\top}\boldsymbol{\beta}-\hat{\boldsymbol{f}}_{i}^{\top}\boldsymbol{\gamma}\right\}^{2}+\lambda\|\boldsymbol{\beta}\|_{1}\right],
	\end{align}
	where $\lambda>0$ is a tuning parameter. 
	
	By utilizing pseudo response observations $F_{n}(Y_{i})$ instead of $Y_{i}$, our approach is inherently robust when confronting with outliers. We note that for the SIM without latent factors, the distribution function transformation has also been considered by many other authors, such as \cite{zhu2009nonconcave}, \cite{wang2012non}, and \cite{rejchel2020rank}. However, in our model $\boldsymbol{f}$ and $\boldsymbol{u}$ are unobserved and must be estimated firstly. Thus we require additional efforts to derive the theoretical properties of $\hat{\boldsymbol{\beta}}_{h}$.
	
	Let $\tilde{F}_{n}(\boldsymbol{Y})=(\boldsymbol{I}_{n}-\hat{\boldsymbol{P}})\left\{F_{n}(\boldsymbol{Y})-1/2\right\}$ represent the residuals of the response vector $F_{n}(\boldsymbol{Y})-1/2$ after it has been projected onto the column space of $\hat{\boldsymbol{F}}$, where $\hat{\boldsymbol{P}}=n^{-1}\hat{\boldsymbol{F}}\hat{\boldsymbol{F}}^{\top}$  is the corresponding projection matrix. Since $\hat{\boldsymbol{U}}=(\boldsymbol{I}_{n}-\hat{\boldsymbol{P}})\boldsymbol{X}$, $\hat{\boldsymbol{F}}^{\top}\hat{\boldsymbol{U}}={\boldsymbol{0}_{K \times p}}$. Then direct calculations yield  that the solution of \eqref{empiricalestimator} is
	equivalent to 
	\begin{align}
		\hat{\boldsymbol{\beta}}_{h}&=\arg\min\limits_{\boldsymbol{\beta}\in \mathbb{R}^{p}} \left\{\frac{1}{2n}\left\|\tilde{F}_{n}(\boldsymbol{Y})-\hat{\boldsymbol{U}}\boldsymbol{\beta}\right\|_{2}^{2}+\lambda\|\boldsymbol{\beta}\|_{1}\right\},  \label{projection form of loss function} \\
		\hat{\boldsymbol{\gamma}}_{h}&=(\hat{\boldsymbol{F}}^{\top}\hat{\boldsymbol{F}})^{-1}\hat{\boldsymbol{F}}^{\top}\left\{F_{n}(\boldsymbol{Y})-\frac{1}{2}\right\}. \label{the least squares estimate of phi}
	\end{align}
	
	For any vector $\hat{\boldsymbol{\beta}}_{h}$ defined in \eqref{projection form of loss function}, we have the following result.
	\begin{theorem} \label{consistency property}
		Under Assumptions \ref{factorassumption}-\ref{Loadings and initial pilot estimators} and LC, assume that $\log p=o(n)$. If  $s=o(n/\log p)$ and $\lambda \asymp \sqrt{(\log p)/n}$, then $\hat{\boldsymbol{\beta}}_{h}$ defined in  \eqref{projection form of loss function} satisfies
		\begin{align*}
			\|\hat{\boldsymbol{\beta}}_{h}-\boldsymbol{\beta}_{h}\|_{2}=O_{\mathbb{P}}(\lambda \sqrt{s}), \ \ and \ \ \|\hat{\boldsymbol{\beta}}_{h}-\boldsymbol{\beta}_{h}\|_{1}=O_{\mathbb{P}}(\lambda s). 
		\end{align*}
		%with probability at least $1-c_{1}\exp(-c_{2}log  p)$. 
		
	\end{theorem}
	
	Based on the empirical distribution function of the response, we establish the $\ell_{1}$ and $\ell_{2}$-error bounds for our introduced estimator $\hat{\boldsymbol{\beta}}_{h}$ in Theorem \ref{consistency property}.
	Compared with \cite{zhu2009nonconcave}, \cite{wang2012non}, and \cite{rejchel2020rank}, in our model $\boldsymbol{f}$ and $\boldsymbol{u}$ are unobserved and must be estimated firstly. This adds new technical difficulty. Further compared with \cite{fanJ2023}, our procedures do not need any moment condition for the error term $\varepsilon$ in the model. 
	
	\subsection{Inference of FASIM}
	Due to the bias inherent in the penalized estimation, $\hat{\boldsymbol{\beta}}_{h}$ is unsuitable for direct utilization in statistical inference. To overcome this obstacle, \cite{zhang2014confidence} and \cite{van2014asymptotically} proposed the debiasing technique for Lasso estimator in linear regression. \cite{eftekhari2021inference} extended the technique to SIM. \cite{han2023robust} and \cite{Yang2024Robust} further considered robust inference for high-dimensional SIM. However, their procedures cannot handle latent factors. To this end, we introduce debiased estimator for the FASIM.
	Motivated by \cite{zhang2014confidence} and \cite{van2014asymptotically}, we construct the following debiased estimator 
	\begin{align} \label{betatilde}
		\tilde{\boldsymbol{\beta}}_{h}=\hat{\boldsymbol{\beta}}_{h}+\frac{1}{n}\hat{\boldsymbol{\Theta}}\hat{\boldsymbol{U}}^{\top}\left\{F_{n}(\boldsymbol{Y})-\frac{1}{2}-\hat{\boldsymbol{U}}\hat{\boldsymbol{\beta}}_h\right\},
	\end{align}
	where $\hat{\boldsymbol{\Theta}}$ is  a sparse estimator of $\boldsymbol{\Sigma}_{u}^{-1}$, with $\boldsymbol{\Sigma}_{u}=\mathbb{E}\left({\boldsymbol{u}}_{i}{\boldsymbol{u}}_{i}^{\top}\right)$.
	Denote $\hat{\boldsymbol\Sigma}_{u}=n^{-1}\sum_{i=1}^{n}\hat{\boldsymbol{u}}_{i}\hat{\boldsymbol{u}}_{i}^{\top}$. Then we construct ${\hat{\boldsymbol{\Theta}}}$ by employing the approach in \cite{cai2011constrained}. Concretely,  ${\hat{\boldsymbol{\Theta}}}$ is the solution to the following
	constrained optimization problem: 
	\begin{align}\label{Theta hat estimation}
		\hat{\boldsymbol{\Theta}}&=\arg\min\limits_{\boldsymbol{\Theta} \in \mathbb{R}^{p \times p}}\|\boldsymbol{\Theta}\|_{\text{sum}}, \notag \\
		&\text{s.t.} \ \|\boldsymbol{\Theta}\hat{\boldsymbol{\Sigma}}_{u}-\boldsymbol{I}_{p}\|_{\max}\leq \delta_{n},
	\end{align}
	where $\delta_{n}$ is a predetermined tuning parameter.
	In general, $\hat{\boldsymbol{\Theta}}$ is not symmetric since there is no symmetry constraint in \eqref{Theta hat estimation}. Symmetry can be enforced through additional operations. Denote $\hat{\boldsymbol{\Theta}}=\left(\hat{\boldsymbol{\xi}}_{1},\ldots,\hat{\boldsymbol{\xi}}_{p}\right)^{\top}=(\hat{\xi}_{ij})_{1 \leq i,j \leq p}$. Write $\hat{\boldsymbol{\Theta}}^{\text{sym}}=\left(\hat{\xi}_{ij}^{\text{sym}}\right)_{1 \leq i,j \leq p}$, where $\hat{\xi}_{ij}^{\text{sym}}$ is defined as:
	\begin{align}
		\hat{\xi}_{ij}^{\text{sym}}=\hat{\xi}_{ij}\mathbb{I}\left(|\hat{\xi}_{ij}| \leq |\hat{\xi}_{ji}|\right)+\hat{\xi}_{ji}\mathbb{I}\left(|\hat{\xi}_{ij}|> |\hat{\xi}_{ji}|\right). \notag 
	\end{align}
	Apparently, $\hat{\boldsymbol{\Theta}}^{\text{sym}}$ is a symmetric matrix. For simplicity, we write $\hat{\boldsymbol{\Theta}}$ as the symmetric estimator in the rest of the paper. 
	Next, we consider the estimation error bound of $\hat{\boldsymbol{\Theta}}$.  To achieve this target, we need to introduce the following assumption on the inverse of the $\boldsymbol{\Sigma}_{u}$.
	\begin{assumption}\label{assumption on the sparsity of the inverse of SIgmau}
		There exists a positive constant $M^{*}$ such that $\left\|\boldsymbol{\Sigma}_{u}^{-1}\right\|_{\infty} \leq M^{*}$. Moreover,  $\boldsymbol{\Sigma}_{u}^{-1}=\left({\boldsymbol{\xi}}_{1},\ldots,{\boldsymbol{\xi}}_{p}\right)^{\top}=({\xi}_{ij})_{1 \leq i,j \leq p}$ is row-wise sparse, i.e., $\max_{i \in [p]}\sum_{j=1}^{p}|{\xi}_{ij}|^{q} \leq c_{n,p}$,  where $c_{n,p}$ is positive and bounded away from zero and allowed to increase as $n$ and $p$ grow, and $0 \leq q < 1$.
	\end{assumption}
	
	Assumption \ref{assumption on the sparsity of the inverse of SIgmau} necessitates that  $\boldsymbol{\Sigma}_{u}^{-1}$  be sparse in terms of both its $\ell_{\infty}$-norm and matrix row space. \cite{van2014asymptotically}, \cite{cai2016estimating} and \cite{ning2017general} also discussed similar assumptions  on precision matrix estimation
	and more general inverse Hessian matrix estimation.
	
	The estimation error bound of $\hat{\boldsymbol{\Theta}}$ and the upper bound of $\left\|\hat{\boldsymbol{\Theta}}-\boldsymbol{\Sigma}_{u}^{-1}\right\|_{\infty}$ are shown in Proposition 7 and Lemma 8 in the Supplementary Material, which are crucial  for establishing theoretical results afterwards. Denote $\boldsymbol{m}=\{m_{1}(Y),\ldots,m_{p}(Y)\}^{\top}$, with  $m_{j}(y)=\mathbb{E}[(X_{1j}-\boldsymbol{f}_{1}^{\top}\boldsymbol{b}_{j})\{\mathbb{I}(Y\geq y)-F(Y)\}]$. Let $\boldsymbol{m}_i$ be the i.i.d copies of $\boldsymbol{m}$. Namely, $\boldsymbol{m}_i=\{m_{1}(Y_i),\ldots,m_{p}(Y_i)\}^{\top}$.
	
	\begin{theorem} \label{debias decomposition theorem}
		Suppose that the conditions in Theorem \ref{consistency property}  and Assumption \ref{assumption on the sparsity of the inverse of SIgmau} are satisfied. 
		Then if $s=o(\sqrt{n}/\log p)$ and $\log p << n^{{(1-q)}/{(3-q)}}$, where $q$ is given in Assumption \ref{assumption on the sparsity of the inverse of SIgmau},  we have 
		\begin{align}
			&\sqrt{n}(\tilde{\boldsymbol{\beta}}_{h}-{\boldsymbol{\beta}}_{h})=\boldsymbol{Z}_{b}+\boldsymbol{E}, \notag \\
			&\boldsymbol{Z}_{b}=\frac{1}{\sqrt{n}}\sum_{i=1}^n\boldsymbol{\Sigma}_{u}^{-1}\left({\boldsymbol{u}}_ie_{hi}+\boldsymbol{m}_i\right), \ \|\boldsymbol{E}\|_{\infty}=o_{p}(1). \notag 
		\end{align}
	\end{theorem}
	
	Theorem \ref{debias decomposition theorem} indicates that the asymptotic representation of $\sqrt{n}(\tilde{\boldsymbol{\beta}}_{h}-{\boldsymbol{\beta}}_{h})$ can be divided into two terms. The major term $\boldsymbol{Z}_{b}$ is associated with the error $\boldsymbol{e}_{h}$ and $\boldsymbol{m}$, while the remainder $\boldsymbol{E}$ vanishes as $n$  approaches to infinity. 
	
	Further for $j =1,\ldots,p$, the  variance  of ${Z}_{bj}$ is ${\sigma}_{zj}^{2}=\boldsymbol{e}_j^\top\boldsymbol{\Sigma}_{u}^{-1}
	\mbox{Var}\{\boldsymbol{u}e_{h}+\boldsymbol{m}\}
	\boldsymbol{\Sigma}_{u}^{-1}\boldsymbol{e}_j$,
	with $\boldsymbol{e}_j$ being an unit vector with only its $j$-th element being 1. %and ${m}_{j}(y)=\mathbb{E}[(X_{1j}-\boldsymbol{f}_{1}^{\top}\boldsymbol{b}_{j})\{\mathbb{I}(Y\geq y)-F(Y)\}]$ defined in \eqref{decorrelated score function}. 
	Based on Theorem \ref{debias decomposition theorem}, the asymptotic variance of $\sqrt{n}(\tilde{{\beta}}_{hj}-{{\beta}}_{hj})$ can be estimated by {$\hat{\sigma}_{zj}^{2}=\boldsymbol{e}_j^\top\hat{\boldsymbol{\Theta}}\tilde{\boldsymbol{\Sigma}}\hat{\boldsymbol{\Theta}}\boldsymbol{e}_j$, with $\tilde{\boldsymbol{\Sigma}}=(\tilde{\sigma}_{lk})_{1\leq l,k \leq p}$, $\tilde{\sigma}_{lk}={n}^{-1}\sum_{i=1}^{n}\left\{\hat{U}_{il}\tilde{e}_{hi}+\hat{m}_{l}(Y_{i})\right\}\left\{\hat{U}_{ik}\tilde{e}_{hi}+\hat{m}_{k}(Y_{i})\right\}$, $\tilde{e}_{hi}=F_{n}(Y_{i})-{1}/{2}-\hat{\boldsymbol{u}}_{i}^{\top}\hat{\boldsymbol{\beta}}_{h}-\hat{\boldsymbol{f}}_{i}^{\top}\hat{\boldsymbol{\gamma}}_{h}$ and $\hat{m}_{j}(y)={n}^{-1}\sum_{i=1}^{n}\left(X_{ij}-\hat{\boldsymbol{f}}_{i}^{\top}\hat{\boldsymbol{b}}_{j}\right)\{\mathbb{I}(Y_{i}\geq y)-F_{n}(Y_{i})\}$.
		Thus the $(1-\alpha)$ confidence interval for $\beta_{hj}$, $j \in [p]$ can be constructed as follows
		\begin{align} \label{confidence interval}
			\mathcal{CI}_{\alpha}(\beta_{hj})=\left(\tilde{\beta}_{hj}-\frac{\hat{\sigma}_{zj}z_{1-\alpha/2}}{\sqrt{n}}, \  \tilde{\beta}_{hj}+\frac{\hat{\sigma}_{zj}z_{1-\alpha/2}}{\sqrt{n}}\right),
		\end{align}
		where $\tilde{\beta}_{hj}$ is the $j$-th component of $\tilde{\boldsymbol{\beta}}_{h}$ and $z_{1-\alpha/2}$ is the $(1-\alpha/2)$-th quantile of standard normal distribution.}
	
	\section{Numerical studies} \label{Numerical studies section}
	In this section,  simulation studies are conducted to evaluate the finite sample performance of the proposed methods.  We implement the proposed method with the following two models.\\
	\textbf{Model 1}: Linear model 
	\begin{align} \label{simulation generate data linear model}
		Y=\boldsymbol{u}^{\top}\boldsymbol{\beta}+\boldsymbol{f}^{\top}\boldsymbol{\gamma}+\varepsilon. 
	\end{align}
	\textbf{Model 2}: Nonlinear model 
	\begin{align} \label{simulation generate data Non-linear model}
		Y=\exp\left(\boldsymbol{u}^{\top}\boldsymbol{\beta}+\boldsymbol{f}^{\top}\boldsymbol{\gamma}+\varepsilon \right).
	\end{align} 
	Under each setting, we generate $n$ i.i.d. observations, and replicate 500 times.
	
	\subsection{Adequacy test of factor model}
	In this subsection,  we set $n=200$, $K=2$, $p=200$ or $500$, and ${\bX}=\boldsymbol{F}\boldsymbol{B}^{\top}+\boldsymbol{U}$.
	Here, the entries of $\boldsymbol{B}$ are  generated from the uniform  distribution $\mathrm{Unif}(-1,1)$ and every row of $\boldsymbol{U}$ follows from $\mathrm{N}(\boldsymbol{0},\boldsymbol{\Sigma}_{\boldsymbol{u}})$ with $\boldsymbol{\Sigma}_{\boldsymbol{u}}=(\sigma_{u ij})_{1\leq i,j \leq p}$ and  $\sigma_{uij}=0.5^{|i-j|}$.
	We consider two cases of $\boldsymbol{F}$ generation:\\
	Case i. We generate $\boldsymbol{f}_{i}$ from standard normal distribution, that is, 
	\begin{align}
		\boldsymbol{f}_{i}\sim \mathrm{N}(\boldsymbol{0},\boldsymbol{I}_{K}), \ i=1,\ldots,n. \label{generate F model1}
	\end{align}
	Case ii. We generate  $\boldsymbol{f}_{i}$  from the AR(1) model:
	\begin{align}
		\boldsymbol{f}_{i}=\boldsymbol{\Phi}\boldsymbol{f}_{i-1}+\boldsymbol{\pi}_{i}, \, i=2,\ldots,n,\label{generate F model2}
	\end{align}
	where $\boldsymbol{\Phi}=(\phi_{ij}) \in \mathbb{R}^{K\times K}$ with ${\phi}_{ij}=0.4^{|i-j|+1}, \ i,j \in [K]$. In addition, ${\boldsymbol{f}_{1}}$ and ${\boldsymbol{\pi}_{i}}, \ i \geq 1$ are  independently drawn from $\mathrm{N}(\boldsymbol{0},\boldsymbol{I}_{K})$. We generate ${\varepsilon}$ either from (a) $\mathrm{N}(0,0.25)$ or  (b) Student’s t distribution with 3 degree of freedom, denoted as $\mathrm{t}_3$.
	We set  $\gamma=(0.5, 0.5)^{\top}$ and $\boldsymbol{\beta}=\omega\ast\left(\mathbf{1}_{3},\mathbf{0}_{p-3}\right)^{\top}$, $\omega \geq 0$.
	When $\omega=0$, it indicates that the null hypothesis holds, and the simulation results correspond to the empirical size. Otherwise, they correspond to the empirical power.
	
	To assess the robustness of our proposed method,  we introduce outliers to  the responses. 
	We randomly pick $p_{out}$ of the response, and increase by $m_{out}$-times maximum of original responses, shorted as $p_{out}+m_{out}*\max$(response).
	Here, $p_{out}$ represents the proportion of outliers, while $m_{out}$ is a predetermined constant indicating the strength of the outliers.
	Throughout the simulation, we adopt the strategy of 10\%+10$*\max$(response) to pollute the observations. 
	We compare our proposed FAST with FabTest in \cite{fanJ2023}.   This comparison under the linear model \eqref{simulation generate data linear model}  is illustrated in Figure \ref{Linear_P_200_our_Fan_figure} and Figure \ref{Linear_P_500_our_Fan_figure}. Here, ``FAST\_i'' and ``FAST\_ii'' signify the results derived from the FAST  corresponding to settings \eqref{generate F model1} and \eqref{generate F model2} of $\boldsymbol{F}$ generation, respectively. Similarly, ``FabTest\_i'' and ``FabTest\_ii'' denote the results of FabTest in \cite{fanJ2023} corresponding to settings \eqref{generate F model1} and \eqref{generate F model2} of $\boldsymbol{F}$ generation, respectively. The first rows in Figures \ref{Linear_P_200_our_Fan_figure} and \ref{Linear_P_500_our_Fan_figure} represent the results obtained with the original data, while the second rows correspond to the results obtained after incorporating outliers (namely, 10\%+10$*\max$(response)). The figures suggest that both FAST and FabTest  exhibit commendable performance under the Gaussian distribution. But when confronted  with heavy tails or outliers, our proposed method  outperforms FabTest. Specifically, as illustrated in the second rows of these two figures,  the power curves associated with FAST demonstrate a notably swifter attainment of 1. In contrast, the power curves related to FabTest consistently hover around the significance level of 0.05.
	
	\begin{center}
		Figures \ref{Linear_P_200_our_Fan_figure} and \ref{Linear_P_500_our_Fan_figure} should be here.
	\end{center}

	Figures \ref{Non_Linear_P_200_our_Fan_figure} and \ref{Non_Linear_P_500_our_Fan_figure} illustrate the power curves of FAST and FabTest under the nonlinear model \eqref{simulation generate data Non-linear model}.
	We can observe a distinct performance in the nonlinear scenario where, even with the light-tailed error distribution {$\mathrm{N}(0, 0.25)$} within the original data, the power performance of the FabTest is notably inferior compared to that of FAST. Furthermore, when considering the $\mathrm{t}_3$-distribution within the original data, the power curves of FAST continue to reach 1 faster. Similarly, when outliers are introduced, FabTest exhibits a complete failure, whereas FAST continues to perform very well. 
	These results serve to illustrate the robustness of FAST in  testing.
	
	\begin{center}
		Figures \ref{Non_Linear_P_200_our_Fan_figure} and \ref{Non_Linear_P_500_our_Fan_figure} should be here.
	\end{center}

	As previously mentioned, our method avoids estimating high-dimensional regression coefficients and high-dimensional precision matrix, significantly reducing computational costs. Table \ref{computation time} provides a comparison of {the average computation time} between the FAST and the FabTest in \cite{fanJ2023} under the linear model \eqref{simulation generate data linear model}.  The table indicates  that under the same settings, the average computation time for FAST is significantly less than that for FabTest. Additionally, it is evident that the average computation time for both tests increase as the parameter dimension $p$ increases to 500, which is expected.
	
	\begin{center}
		Table \ref{computation time} should be here.
	\end{center}

	\subsection{Accuracy of estimation} \label{Accuracy of estimation}
	To illustrate the accuracy of our estimation,
	we set  the number of latent factors to  $K=2$,   the dimension of $\boldsymbol{x}$ to $p=500$, $\gamma=(0.5, 0.5)^{\top}$, the first $s=3$ entries of $\boldsymbol{\beta}$ to 0.5, the remaining $p-s$ entries to 0.
	Throughout this subsection, we generate each entry of $\boldsymbol{F}$ and $\boldsymbol{U}$ from the standard Gaussian distribution $\mathrm{N}(0, 1)$, while each entry of $\boldsymbol{B}$ is randomly generated from the uniform distribution $\mathrm{Unif}(-1, 1)$.  The
	$\varepsilon$ is generated 
	from the three scenarios:  (a) the standard Gaussian distribution $\mathrm{N}(0, 1)$; (b) the uniform distribution $\mathrm{Unif}(-{3}/{2}, {3}/{2})$; (c) $\mathrm{t}_3$.
	Under each setting,  $n$ is set to satisfy that $\sqrt{s(\log p)/n}$ takes uniform grids in $[0.10, 0.30]$ with step being 0.05. 
	
	To validate the necessities of considering latent factors in observational studies, we compare the relative error obtained by our approach (denoted as ``FASIM\_Lasso'') with the method in \cite{rejchel2020rank} (denoted as ``SIM\_Lasso''), which considers the single-index model without incorporating the factor effect,	 {and  the method in \cite{fanJ2023} (denoted as  ``FA\_Lasso'').}
	Figures \ref{l1_linear_Nonlinear_Gaussian_Uniform_figure} and \ref{l2_linear_Nonlinear_Gaussian_Uniform_figure} depict the relative error  $$L_{2}(\hat{\boldsymbol{\beta}}_{h}, \boldsymbol{\beta}_{h})=\frac{\|\hat{\boldsymbol{\beta}}_{h}-\boldsymbol{\beta}_{h}\|_{2}}{\left\|\boldsymbol{\beta}_{h}\right\|_{2}}$$ of linear model \eqref{simulation generate data linear model}  and nonlinear model \eqref{simulation generate data Non-linear model}, respectively. The simulation results indicate that for original data, FA\_Lasso performs very well when the error is not heavy-tailed, but when the error follows the $\mathrm{t}_3$, FA\_Lasso performs significantly worse. In contrast, the proposed method maintains stable and good performance across different error distribution scenarios, illustrating the robustness of our approach. Additionally, compared to SIM\_Lasso,  FASIM\_Lasso  enjoys smaller relative errors. This is partially due to the inadequacy of the SIM in \cite{rejchel2020rank}. Additionally,  Figures \ref{l1_linear_Nonlinear_Gaussian_Uniform_figure} and \ref{l2_linear_Nonlinear_Gaussian_Uniform_figure} show that the upper limits of the statistical rates $\|\hat{\boldsymbol{\beta}}_{h}-\boldsymbol{\beta}_{h}\|_{2}$ of FASIM\_Lasso  are $O\left\{\sqrt{s (\log p)/n}\right\}$, which validates Theorem \ref{consistency property}.
	
	\begin{center}
		Figures  \ref{l1_linear_Nonlinear_Gaussian_Uniform_figure} and \ref{l2_linear_Nonlinear_Gaussian_Uniform_figure} should be here.
	\end{center}

	To investigate the robustness of FASIM\_Lasso, 
	we introduce outliers into the observations using the aforementioned method, specifically, 10\%+10$*\max$(response). We output the $L_{2}(\hat{\boldsymbol{\beta}}, \boldsymbol{\beta})$ of FA\_Lasso and $L_{2}(\hat{\boldsymbol{\beta}}_{h}, \boldsymbol{\beta}_{h})$ of FASIM\_Lasso and SIM\_Lasso.
	The results are depicted in Figures \ref{l1_ori_out__Linear_Nonlinear_Gaussian_Uniform_figure} and \ref{l2_ori_out__Linear_Nonlinear_Gaussian_Uniform_figure}. The figures indicate that our proposed method can achieve more precise results, with  smaller relative errors in all scenarios. 
	This implies that our proposed method   is much more robust than FA\_Lasso confronted with outliers in the observed data. 
	
	\begin{center}
		Figures \ref{l1_ori_out__Linear_Nonlinear_Gaussian_Uniform_figure} and \ref{l2_ori_out__Linear_Nonlinear_Gaussian_Uniform_figure} should be here.
	\end{center}

	\subsection{Validity of confidence interval}
	In this subsection, we construct confidence intervals based on the proposed method in Section \ref{Estimation and debiased Lasso inference}. We generate ${\varepsilon}$ either from (a) $\mathrm{N}(0,0.25)$ or  (b) $\mathrm{t}_3$ and set $n=200$, $p=200$ or $500$. In addition,   we generate $\boldsymbol{F}$, $\boldsymbol{U}$, $\boldsymbol{B}$ and the parameters $\boldsymbol{\beta}$, $\boldsymbol{\gamma}$ as described   in Section \ref{Accuracy of estimation}. %However,    we consider the sparsity of $\boldsymbol{\beta}$ here to be 20 and   $n=200$. That is, we set the first $s=20$ entries of $\boldsymbol{\beta}$ to 0.5, the remaining $p-s$ entries to 0. 
	To assess the performance of our methods, we  examine the empirical coverage probabilities and the average lengths of confidence intervals for the individual coefficients across all covariates,  covariates on    $\mathcal{S}=\{1,2,3\}$ and  those on $\mathcal{S}^{C}=[p]/\mathcal{S}$. We define 
	\begin{align}
		\text{CP}&=\frac{\sum\limits_{j=1}^{p}\mathbb{I}\{{\beta}_{hj} \in \mathcal{CI}_{\alpha}(\beta_{hj}) \}}{p}, \ \ \ \text{AL}=\frac{\sum\limits_{j=1}^{p}2{\hat{\sigma}_{zj}z_{1-\alpha/2}}/{\sqrt{n}}}{p}, \notag \\
		\text{CP}_{\mathcal{S}}&=\frac{\sum\limits_{j\in \mathcal{S}}\mathbb{I}\{{\beta}_{hj} \in \mathcal{CI}_{\alpha}(\beta_{hj}) \}}{s}, \ \ \ \text{AL}_{\mathcal{S}}=\frac{\sum\limits_{j\in \mathcal{S}}2{\hat{\sigma}_{zj}z_{1-\alpha/2}}/{\sqrt{n}}}{s}, \notag \\
		\text{CP}_{\mathcal{S}^{C}}&=\frac{\sum\limits_{j\in \mathcal{S}^{C}}\mathbb{I}\{{\beta}_{hj} \in \mathcal{CI}_{\alpha}(\beta_{hj}) \}}{p-s}, \ \ \ \text{AL}_{\mathcal{S}^{C}}=\frac{\sum\limits_{j\in \mathcal{S}^{C}}2{\hat{\sigma}_{zj}z_{1-\alpha/2}}/{\sqrt{n}}}{(p-s)}. \notag 
	\end{align}
	The results are summarized in Table \ref{The coverage probability and average length of the confidence intervals.}.  The empirical coverage probabilities  are close to the nominal level across all settings,  and the average lengths of the confidence interval are very short. This illustrates the merits of our proposed methods. In addition, we find that even when the error follows a $\mathrm{t}_3$ distribution, the average interval lengths only increase slightly.
	
	\begin{center}
		Table \ref{The coverage probability and average length of the confidence intervals.} should be here.
	\end{center}

	\section{Real data analysis} \label{Real data analysis section}
	In this section, we employ our method with a macroeconomic dataset named FRED-MD \citep{Mccracken2016fred} which comprises 126 monthly U.S. macroeconomic variables.
	Due to these variables measuring certain aspects of economic health, they are influenced by latent factors and thus can be regarded as  intercorrelated.
	We analyze this dataset to illustrate the performance of FASIM and evaluate the adequacy of the factor regression model. 
	In our study, we take ``GS5''  as response variable $Y$, and the remaining variables as predictors $\boldsymbol{x}$, where  ``GS5'' represents  the 5-year treasury rate.		
	In addition, to demonstrate the robustness of FASIM, we apply the previously mentioned outlier handling method to the response variable, specifically using 10\%+10$*\max$(response). 
	Due to the economic crisis of 2008-2009, the data remained unstable even after implementing the suggested transformations.
	Therefore, we examine data from two distinct periods: February 1992 to October 2007 and August 2010 to February 2020.
	
	At the beginning,  we employ our introduced FAST and FabTest in \cite{fanJ2023} to test the adequacy of factor models, and  denote the corresponding models with only factors as F\_SIM and F\_LM, respectively. {We employ 2000 bootstrap replications here to obtain the critical value.  For  FabTest, we  use 10 fold cross-validation to compute the tuning parameters and  refit cross-validation based on iterated sure independent screening to estimate the variance of the error.} 
	The $p$-values  are provided in Table \ref{$p$-values of each test for the real data based on FASIM  and the FARM.}.
	At the significance level of 0.05,  { both for the original dataset and the polluted dataset, the results indicate the inadequacy of F\_SIM for the  ``GS5'' within the two distinct time periods.
		While for F\_LM,  the results indicate that it is inadequate for ``GS5''  in  the original  dataset  within the two distinct time periods. When the data is polluted, the null hypothesis is rejected  during the period from February 1992 to October 2007, while it is not rejected  during the other period.}
	%But during July 2010 to October 2023, the results indicate that F\_SIM is inadequate for ``GS5''  under both the original and  polluted dataset scenarios, while  F\_LM  is inadequate for the original   dataset but adequate for polluted  dataset. 
	This implies the necessary to introduce the idiosyncratic component 
	$\boldsymbol{u}$ into the regression model. Hence, in the subsequent study on prediction accuracy, we consider FASIM and FARM for comparison.
	
	\begin{center}
		Table \ref{$p$-values of each test for the real data based on FASIM  and the FARM.} should be here.
	\end{center}
	
	We compare the forecasting results of FASIM with that of FARM. For each given time period and model, predictions are performed using a moving window approach with a window size of 90 months. Indexing the panel data from 1 for each of the two time periods, for all \( t > 90 \), we use the 90 previous measurements \(\{(\boldsymbol{x}_{t-90}, Y_{t-90}), \ldots, (\boldsymbol{x}_{t-1}, Y_{t-1})\}\) to train the models (FASIM and FARM) and output predictions \(\hat{Y}_{\text{FASIM},t}\) and \(\hat{Y}_{\text{FARM},t}\). For FASIM, as defined in Equation \eqref{FASIM}, after obtaining the estimated parameters \(\hat{\boldsymbol{\beta}}_{h}\) and \(\hat{\boldsymbol{\gamma}}_{h}\), the estimated latent factor vector \(\hat{\boldsymbol{f}}_{t}\), and the estimated  idiosyncratic component \(\hat{\boldsymbol{u}}_{t}\) at the \( t \)-th time point, we get the estimator \(\hat{g}\) of the unknown function \( g \) via spline regression. Finally, the predicted value of the response variable is calculated as \(\hat{Y}_{t}=\hat{g}(\hat{\boldsymbol{f}}_{t}^{\top}\hat{\boldsymbol{\gamma}}+\hat{\boldsymbol{u}}_{t}^{\top}\hat{\boldsymbol{\beta}})\). The accuracy of FASIM and FARM is measured using the Mean Square Error (MSE) \citep{hastie2009elements}, defined as:
	\[
	\text{MSE} = \frac{1}{T-90} \sum_{t=91}^{T} (Y_{t} - \hat{Y}_{t})^{2},
	\]
	where \( T \) denotes the total number of data points in a given time period.

	\begin{center}
		Table \ref{Accuracy of  FASIM, FARM and SIM based on real data.} should be here.
	\end{center}

	Table \ref{Accuracy of  FASIM, FARM and SIM based on real data.} presents the prediction accuracy results of FASIM and FARM of  ``GS5'' for the original dataset and the polluted  dataset within the two distinct time periods.
	It is evident that  the performance of FASIM and FARM on the original data are similar. However, while the MSEs of both FASIM and FARM increase for the polluted dataset, the increase is substantial for FARM, whereas FASIM shows only a modest increase. This suggests that FASIM is more robust than FARM.

	\section{Conclusions and discussions} \label{Conclusions and discussions section}
	To capture nonliearity with latent factors, in this paper, we introduce a novel Factor Augmented sparse Single-Index Model, FASIM. For this newly {proposed} model, we first address the concern  of whether the augmented part is necessary or not. We develop a score-type test statistic without estimating high-dimensional regression coefficients nor high-dimensional precision matrix. We also propose a Gaussian multiplier bootstrap to determine the critical value for our proposed test statistic FAST. The validity of our procedure is theoretically established under {mild}
	conditions. When the model test is passed, we employ the $\ell_1$ penalty to estimate the unknown parameters, establishing both \(\ell_{1}\) and \(\ell_{2}\) error bounds. We also introduce debiased estimator and establish its asymptotic normality. Numerical results illustrate the robustness and effectiveness of our proposed procedures.
	
	In practice, it would be of interest to test whether the FASIM is actually FARM. We may also consider the multiple testing problem for the FASIM. Further the condition $\kappa_h \neq 0$ excludes even link functions, and in particular the problem of sparse phase retrieval. We would explore these possible topics in near future. 
	\newpage
	\section*{Appendix}
	\appendix
	
	\label{app}
	\allowdisplaybreaks[3]
	
	\renewcommand{\theequation}{\thesection.\arabic{equation}}
	
	\setcounter{table}{0}
	\renewcommand{\thetable}{\thesection\arabic{table}}
	\setcounter{figure}{0}
	\renewcommand{\thefigure}{\thesection\arabic{figure}}
	\section{Proofs of Theorems} \label{Proofs of Theorems}
	In the following, we will present the proofs of our main theorems. To save space, proofs of  some technical lemmas will be 	shown in Supplementary Material.
	\subsection{Proof of Theorem \ref{Score Gaussianapproximationtheoremresult} } \label{Score Gaussianapproximationtheoremresult proof}
	\begin{proof}
		By Lemma 2 in Supplementary Material, with high probability, we have 
		\begin{align}
			\left|\max\limits_{j \in [p]}|T_{nj}|-	\max\limits_{j \in [p]}|S_{nj}|\right| \leq \max\limits_{j \in [p]} \left|T_{nj}-S_{nj}\right| \leq Cr(n,p), \notag
		\end{align}
		for some constant $C$,  as $(n,p) \rightarrow \infty$. Here $r(n,p)=O\left\{{(\log p)^{3/2}\log (np)}/{n}+{(\log p)(\log n)}/{\sqrt{n}}\right\}$. This implies that 
		\begin{align} \label{Gaussian Approximation initial decomposition1}
			& \quad \left|\Pr\left(\max\limits_{j \in [p]}|T_{nj}|\leq t\right)-\Pr\left(\|\boldsymbol{\mathrm{Z}}\|_{\infty}\leq t\right)\right| \notag \\
			& \leq \underbrace{\left|\Pr\left\{\max\limits_{j \in [p]}|S_{nj	}| \leq t+Cr(n,p)\right\}
				-\Pr\left\{\|\boldsymbol{\mathrm{Z}}\|_{\infty}\leq t+Cr(n,p)\right\}\right|}_{\text{I}_{\text{B},1}} \notag \\
			& \quad +\underbrace{\left|\Pr\left\{\|\boldsymbol{\mathrm{Z}}\|_{\infty}\leq t+Cr(n,p)\right\}-\Pr\left(\|\boldsymbol{\mathrm{Z}}\|_{\infty}\leq t\right)\right|}_{\text{I}_{\text{B},2}}.
		\end{align}
		For the term $\text{I}_{\text{B},1}=\left|\Pr\left\{\max_{j \in [p]}|S_{nj	}| \leq t+Cr(n,p)\right\}
		-\Pr\left\{\|\boldsymbol{\mathrm{Z}}\|_{\infty}\leq t+Cr(n,p)\right\}\right|$ in \eqref{Gaussian Approximation initial decomposition1}, recall that $S_{nj}={n}^{-1/2}\sum_{i=1}^{n}\left[\left\{F(Y_{i})-{1}/{2}-\boldsymbol{f}_{i}^{\top}\boldsymbol{\gamma}_{h}\right\}\left(X_{ij}-\boldsymbol{f}_{i}^{\top}\boldsymbol{b}_{j}\right)+m_{j}(Y_{i})\right]$  with $m_{j}(y)=\mathbb{E}[(X_{1j}-\boldsymbol{f}_{1}^{\top}\boldsymbol{b}_{j})\{\mathbb{I}(Y\geq y)-F(Y)\}]$, by the sub-Gaussian assumptions of $\boldsymbol{f}$ and $\boldsymbol{u}$, we can get that $\left[\left\{F(Y_{i})-{1}/{2}-\boldsymbol{f}_{i}^{\top}\boldsymbol{\gamma}_{h}\right\}\left(X_{ij}-\boldsymbol{f}_{i}^{\top}\boldsymbol{b}_{j}\right)+m_{j}(Y_{i})\right]_{i=1}^{n}$ is a sub-Exponential variable sequence with bounded norm. In addition, by Assumption \ref{Assumption score}, applying Lemma 3 in Supplementary Material, we have 
		\begin{align} \label{IB1 order score}
			\text{I}_{\text{B},1}=\lim\limits_{n \rightarrow \infty}\sup \limits_{t \in \mathbb{R}}\left|\Pr\left\{\max\limits_{j \in [p]}|S_{nj	}| \leq t+Cr(n,p)\right\}
			-\Pr\left\{\|\boldsymbol{\mathrm{Z}}\|_{\infty}\leq t+Cr(n,p)\right\}\right|=0.
		\end{align}
		For the term $\text{I}_{\text{B},2}=\left|\Pr\left\{\|\boldsymbol{\mathrm{Z}}\|_{\infty}\leq t+Cr(n,p)\right\}-\Pr\left(\|\boldsymbol{\mathrm{Z}}\|_{\infty}\leq t\right)\right|$ in \eqref{Gaussian Approximation initial decomposition1}, by Lemma 4 in Supplementary Material,  we have
		\begin{align} \label{IB2 order score}
			\text{I}_{\text{B},2}&=\left|\Pr\left\{\|\boldsymbol{\mathrm{Z}}\|_{\infty}\leq t+Cr(n,p)\right\}-\Pr\left(\|\boldsymbol{\mathrm{Z}}\|_{\infty}\leq t\right)\right| \notag \\
			& \lesssim r(n,p)\sqrt{1 \vee \log \left\{\frac{p}{r(n,p)}\right\}} \notag \\
			%& \lesssim r(n,p)\sqrt{\log p} \notag \\
			& \rightarrow 0.
		\end{align}
		Combining \eqref{IB1 order score} with \eqref{IB2 order score}, we have 
		\begin{align}
			\left|\Pr\left(\max\limits_{j \in [p]}|T_{nj}|\leq t\right)-\Pr\left(\|\boldsymbol{\mathrm{Z}}\|_{\infty}\leq t\right)\right| \rightarrow 0. \notag
		\end{align}
		The proof is completed.
	\end{proof}
	
	\subsection{Proof of Theorem \ref{the validity of the  bootstrap procedure score} } \label{Proof of the validity of the  bootstrap procedure score}
	\begin{proof}
		Firstly, we have the following decomposition.
		\begin{align}
			& \quad \left|\Pr\left(M_{n}\leq x\right)-\Pr^{*}(\hat{\mathcal{G}} \leq x)\right| \notag \\
			& =\left|\Pr\left(M_{n}\leq x\right)-\Pr\left(\left\|\boldsymbol{\mathrm{Z}}\right\|_{\infty} \leq x\right)+\Pr\left(\left\|\boldsymbol{\mathrm{Z}}\right\|_{\infty} \leq x\right)-\Pr^{*}(\hat{\mathcal{G}} \leq x)\right| \notag \\
			& \leq \left|\Pr\left(M_{n}\leq x\right)-\Pr\left(\left\|\boldsymbol{\mathrm{Z}}\right\|_{\infty} \leq x\right)\right|
			+\left|\Pr\left(\left\|\boldsymbol{\mathrm{Z}}\right\|_{\infty} \leq x\right)-\Pr^{*}(\hat{\mathcal{G}} \leq x)\right|. \notag 
		\end{align}
		By Theorem \ref{Score Gaussianapproximationtheoremresult}, we have 
		\begin{align}
			\left|\Pr\left(M_{n}\leq x\right)-\Pr\left(\left\|\boldsymbol{\mathrm{Z}}\right\|_{\infty} \leq x\right)\right| \rightarrow 0. \notag 
		\end{align}
		We write 
		\begin{align} \label{rho star}
			\rho^{*}=\sup\limits_{x >0}\left|\Pr\left(\left\|\boldsymbol{\mathrm{Z}}\right\|_{\infty} \leq x\right)-\Pr^{*}(\hat{\mathcal{G}} \leq x)\right|.
		\end{align}
		Denote $\tilde{S}_{nj}=n^{-1/2}\sum_{i=1}^{n}\left[\left\{F_{n}(Y_{i})-{1}/{2}-\hat{\boldsymbol{f}}_{i}^{\top}\hat{\boldsymbol{\gamma}}_{h}\right\}\hat{U}_{ij}+\hat{m}_{j}(Y_{i})\right]\mathrm{N}_{i}$.
		Given  the dataset  $\mathcal{D}=\{\boldsymbol{x}_{1},\ldots,\boldsymbol{x}_{n}, Y_{1},\ldots,Y_{n}\}$,   the covariance matrix of  $\tilde{\boldsymbol{S}}_{n}=(\tilde{S}_{nj})_{j=1}^{p}$ is
		\begin{align}
			\tilde{\boldsymbol{\Omega}}=\Cov(\tilde{\boldsymbol{S}}_{n}|\mathcal{D}), \notag
		\end{align}
		with  $\tilde{{\Omega}}_{jk}=n^{-1}\sum_{i=1}^{n}\{\hat{U}_{ij}\hat{e}_{hi}+\hat{m}_{j}(Y_{i})\}\{\hat{U}_{ik}\hat{e}_{hi}+\hat{m}_{k}(Y_{i})\}$, $j,k =1,\ldots,p$, and    $\hat{e}_{hi}=F_{n}(Y_{i})-{1}/{2}-\hat{\boldsymbol{f}}_{i}^{\top}\hat{\boldsymbol{\gamma}}_{h}$. Recall that $\boldsymbol{\Omega}^{*}=\Cov({\boldsymbol{S}}_{n})$, with  ${\Omega}_{jk}^{*}=\mathbb{E}\left\{{U}_{ij}{e}_{hi}+{m}_{j}(Y_{i})\right\}\left\{{U}_{i k}{e}_{hi}+{m}_{k}(Y_{i})\right\},\ j,k=1,\ldots,p$, and $e_{hi}=F(Y_{i})-{1}/{2}-{\boldsymbol{f}_{i}}^{\top}{\boldsymbol{\gamma}}_{h}$.
		By Lemma 5 in Supplementary Material, we have 
		\begin{align}
			\|\boldsymbol{\Omega}^{*}-	\tilde{\boldsymbol{\Omega}}\|_{\max}=o_{\mathbb{P}}\left(\frac{1}{{\log p}}\right). \notag
		\end{align}
		Then it follows from Lemma 2.1 in \cite{chernozhukov2023nearly} that $
		\rho^{*}\stackrel{\mathbb{P}}{\rightarrow}0$.
		
		Combining this result with Theorem \ref{Score Gaussianapproximationtheoremresult}, we have
		\begin{align}
			\left|\Pr\left(M_{n}\leq x\right)-\Pr^{*}(\hat{\mathcal{G}} \leq x)\right| \rightarrow 0.
		\end{align}
		The proof is completed.	
	\end{proof}
	\subsection{Proof of Theorem \ref{power theorem} } \label{Proof of power theorem}
	\begin{proof}
		Firstly, we have the following decomposition
		\begin{align} \label{power of Tn initial decomposition revise}
			\underbrace{\max\limits_{j \in [p]}\left|\sqrt{n}\beta_{hj}\mu_{j}\right|}_{\text{I}_{\text{H}_{1}}}\leq 	\max\limits_{j \in [p]}|T_{nj}|
			+\underbrace{\max\limits_{j \in [p]}\left|\sqrt{n}\beta_{hj}\mu_{j}-T_{nj}\right|}_{\text{I}_{\text{H}_{2}}}.
		\end{align}
		\iffalse
		Using the inequality $2a_{1}a_{2}\leq \epsilon^{-1}a_{1}^{2}+\epsilon a_{2}^{2}$ for any $\epsilon >0$, we have the following equation:
		\begin{align} \label{power of Tn initial decomposition}
			\underbrace{\max\limits_{j \in [p]}\left(\hat{\sigma}_{j}^{-1}\sqrt{n}\beta_{hj}\mu_{j}\right)^{2}}_{\text{I}_{\text{H}_{1}}}\leq 	\max\limits_{j \in [p]}(1+\epsilon)\hat{\sigma}_{j}^{-2}T_{nj}^{2}
			+\underbrace{(1+\epsilon^{-1})\max\limits_{j \in [p]}\hat{\sigma}_{j}^{-2}\left(\sqrt{n}\beta_{hj}\mu_{j}-T_{nj}\right)^{2}}_{\text{I}_{\text{H}_{2}}}.
		\end{align}
		Here, $\mu_{j}=\mathbb{E}(U_{ij}^{2})$.
		\fi
		For the term $\text{I}_{\text{H}_{2}}=\max_{j \in [p]}\left|\sqrt{n}\beta_{hj}\mu_{j}-T_{nj}\right|$ in \eqref{power of Tn initial decomposition revise}, denoted $\eta^{\natural}=\max_{j \in [p]}|(T_{nj}-\sqrt{n}\beta_{hj}\mu_{j})-\left(S_{nj}-\sqrt{n}\beta_{hj}\mu_{j}\right)|$, we have $\eta^{\natural}=O_{\mathbb{P}}\{r(n,p)\}$, with $r(n,p)=(\log p)^{3/2}\log (np)/{n}+(\log p)(\log n)/{\sqrt{n}}+s\|\boldsymbol{\beta}_{h}\|_{\infty}{\log p}/{\sqrt{n}}+{\max_{j  				\in [p]}|\mu_{j}\beta_{hj}|}/{\sqrt{n}}$ by Lemma 2 in Supplementary Material.
		For simplify, we rewrite  $S_{nj}-\sqrt{n}\beta_{hj}\mu_{j}=\sum_{i=1}^{n}\varpi_{ij}/\sqrt{n}$, where $\varpi_{ij}=\left\{F(Y_{i})-{1}/{2}-\boldsymbol{f}_{i}^{\top}\boldsymbol{\gamma}_{h}\right\}U_{ij}+m_{j}(Y_{i})-\beta_{hj}\mu_{j}$.
		Under Assumption \ref{factorassumption}, it's easy
		to show that $\{\varpi_{ij}\}_{i=1}^{n}$ is an i.i.d.  mean zero  sub-Exponential random variable sequence.
		Assume that $\mu_j=\mathbb{E}(U_{ij}^2) \geq C_{\min}$. 
		Denote $K'=\max_{i \in [n]}\|\varpi_{ij}\|_{\psi_{1}}$, by the  Bernstein's inequality,  we have 
		%				\begin{align}
			%					\Pr\left(\max\limits_{j \in [p]}\frac{1}{\sqrt{n}}\sum_{i=1}^{n}\varpi_{ij}\right) \leq 2p \exp\left\{-C\min\left(\frac{t^{2}}{K}\right)\right\}
			%				\end{align}
		\begin{align} \label{sn-n 0.5 order}
			\max\limits_{j \in [p]}\left|S_{nj}-\sqrt{n}\beta_{hj}\mu_{j}\right|=\max\limits_{j \in [p]}\left|\frac{1}{\sqrt{n}}\sum_{i=1}^{n}\varpi_{ij}\right| \leq C_{\min}\sqrt{\frac{{\log p}}{2}}.
		\end{align}
		with probability at least $1-2\exp\{-C(\log p)/(2K'^{2})\}$.
		Therefore, with high probability, we have 
		\begin{align} \label{Tn part in power}
			\max\limits_{j \in [p]}\left|T_{nj}-\sqrt{n}\beta_{hj}\mu_{j}\right| \leq \max\limits_{j \in [p]}\left|S_{nj}-\sqrt{n}\beta_{hj}\mu_{j}\right|+\eta^{\natural}  \leq C_{\min}\sqrt{\frac{{\log p}}{2}}+Cs\|\boldsymbol{\beta}_{h}\|_{\infty}\frac{\log p}{\sqrt{n}}+\max\limits_{j \in [p]}C\frac{|\mu_{j}\beta_{hj}|}{\sqrt{n}}.
		\end{align}
		For any $\boldsymbol{\beta}_{h} \in \boldsymbol{\Delta}(2+\varrho_{0})$, we have 
		\begin{align} \label{power decomposition2 square order}
			\text{I}_{\text{H}_{1}}=\max\limits_{j \in [p]}\left|\sqrt{n}\beta_{hj}\mu_{j}\right|>C_{\min}\sqrt{(2+\varrho_{0})\log p}.
		\end{align}
		Substituting \eqref{Tn part in power} and \eqref{power decomposition2 square order}  into \eqref{power of Tn initial decomposition revise},   we have 
		\begin{align}
			\max\limits_{j \in [p]}\left|T_{nj}\right|& \geq \max\limits_{j \in [p]}\left|\sqrt{n}\beta_{hj}\mu_{j}\right|-\max\limits_{j \in [p]}\left|\sqrt{n}\beta_{hj}\mu_{j}-T_{nj}\right| \notag \\
			& \geq C_{\min}\sqrt{n}\|\boldsymbol{\beta}_{h}\|_{\infty}-C_{\min}\sqrt{\frac{{\log p}}{2}}-Cs\|\boldsymbol{\beta}_{h}\|_{\infty}\frac{\log p}{\sqrt{n}}-C\max\limits_{j \in [p]}\frac{|\mu_{j}\beta_{hj}|}{\sqrt{n}}. \notag 
		\end{align}
		By Theorem \ref{the validity of the  bootstrap procedure score}, 
		$\hat{c}_{1-\alpha}$ is equal to the $(1-\alpha)$-th quantile of $\max_{j \in [p]}\left|T_{nj}-\sqrt{n}\beta_{hj}\mu_{j}\right|$ asymptotically. By \eqref{Tn part in power}, we have 
		$\hat{c}_{1-\alpha} \leq C_{\min}\sqrt{ \log p/2}+Cs\|\boldsymbol{\beta}_{h}\|_{\infty}{\log p}/{\sqrt{n}}+\max_{j \in [p]}C{|\mu_{j}\beta_{hj}|}/{\sqrt{n}}$  
		with high probability.
		%Denote $C^{*}=C_{\min}-2s(\log p)/{n}$.
		%By choosing $(\varrho_{0}, C,C^{*})$ satisfying that $C^{*2}(2+\varrho_{0}) \geq 4C^{2}$, we have 
		Therefore,
		\begin{align}
			\lim_{(n,p) \rightarrow \infty} \inf \limits_{\boldsymbol{\beta}_{h} \in \boldsymbol{\Delta}(2+\varrho_{0})}\Pr\left(M_{n} \geq \hat{c}_{1-\alpha}\right)=1. \notag 
		\end{align}
		The proof is completed.
	\end{proof}
	
	\subsection{Proof of Theorem \ref{consistency property} }
	\begin{proof}
		Recall that $\hat{\boldsymbol{\beta}}_{h}$ is the minimizer of optimization problem as 		
		\begin{align}	\hat{\boldsymbol{\beta}}_{h}=\arg\min\limits_{\boldsymbol{\beta}\in \mathbb{R}^{p}} \left\{\frac{1}{2n}\left\|\tilde{F}_{n}(\boldsymbol{Y})-\hat{\boldsymbol{U}}\boldsymbol{\beta}\right\|_{2}^{2}+\lambda\|\boldsymbol{\beta}\|_{1}\right\}. \notag
		\end{align}
		For the loss function  $\mathcal{L}_{n}({\boldsymbol{\beta}}_{h})=(2n)^{-1}\|(\boldsymbol{I}_{n}-\hat{\boldsymbol{P}})\left\{F_{n}(\boldsymbol{Y})-{1}/{2}\right\}-\hat{\boldsymbol{U}}\boldsymbol{\beta}_{h}\|_{2}^{2}$ in \eqref{projection form of loss function},	to demonstrate the $\ell_{2}$-norm error bound for $\hat{\boldsymbol{\beta}}_{h}$, we aim to prove the following inequality:
		\begin{align} \label{aim inequality}
			C\|\hat{\boldsymbol{\beta}}_{h}-\boldsymbol{\beta}_{h}\|_{2}^{2} \leq \langle\nabla\mathcal{L}_{n}(\hat{\boldsymbol{\beta}}_{h})- \nabla\mathcal{L}_{n}({\boldsymbol{\beta}}_{h}), \hat{\boldsymbol{\beta}}_{h}-\boldsymbol{\beta}_{h}\rangle \leq C_{1}\lambda \sqrt{s}\|\hat{\boldsymbol{\beta}}_{h}-\boldsymbol{\beta}_{h}\|_{2},
		\end{align}
		where $\nabla\mathcal{L}_{n}({\boldsymbol{\beta}}_{h})$ is the gradient of loss function $\mathcal{L}_{n}({\boldsymbol{\beta}}_{h})$, $s=\|\boldsymbol{\beta}_{h}\|_{0}$ and $\lambda\asymp \sqrt{(\log p)/n}$, with positive constants $C$ and $ C_{1}$.
		
		We firstly derive the upper bound of \eqref{aim inequality}. By KKT condition, there exists a subgradient $\boldsymbol{\kappa} \in \partial\|\hat{\boldsymbol{\beta}}_{h}\|_{1}$, such that $\nabla\mathcal{L}_{n}(\hat{\boldsymbol{\beta}}_{h})+\lambda\boldsymbol{\kappa}=\boldsymbol{0}$.   We then derive 
		\begin{align}\label{the upper bound initial decomposition}
			& \quad \langle\nabla\mathcal{L}_{n}(\hat{\boldsymbol{\beta}}_{h})- \nabla\mathcal{L}_{n}({\boldsymbol{\beta}}_{h}), \hat{\boldsymbol{\beta}}_{h}-\boldsymbol{\beta}_{h}\rangle \notag \\
			&=\nabla\mathcal{L}_{n}(\hat{\boldsymbol{\beta}}_{h})^{\top}(\hat{\boldsymbol{\beta}}_{h}-\boldsymbol{\beta}_{h})-\nabla\mathcal{L}_{n}({\boldsymbol{\beta}}_{h})(\hat{\boldsymbol{\beta}}_{h}-\boldsymbol{\beta}_{h}) \notag \\
			&=\underbrace{-\lambda\boldsymbol{\kappa}^{\top}(\hat{\boldsymbol{\beta}}_{h}-\boldsymbol{\beta}_{h})}_{\text{I}_{1}}
			+\underbrace{\nabla\mathcal{L}_{n}({\boldsymbol{\beta}}_{h})(\boldsymbol{\beta}_{h}-\hat{\boldsymbol{\beta}}_{h})}_{\text{I}_{2}}.
		\end{align}
		Denote $\hat{\boldsymbol{\theta}}_{h}=\hat{\boldsymbol{\beta}}_{h}-\boldsymbol{\beta}_{h}$. For $\text{I}_{1}=-\lambda\boldsymbol{\kappa}^{\top}(\hat{\boldsymbol{\beta}}_{h}-\boldsymbol{\beta}_{h})$ in \eqref{the upper bound initial decomposition}, by the definition of  subgradient, we have 
		\begin{align} \label{I1 initial decomposition}
			\text{I}_{1}&=-\lambda\boldsymbol{\kappa}^{\top}(\hat{\boldsymbol{\beta}}_{h}-\boldsymbol{\beta}_{h}) \notag \\
			&\leq \lambda(\|\boldsymbol{\beta}_{h}\|_{1}-\|\hat{\boldsymbol{\beta}}_{h}\|_{1}) \notag \\
			&=\lambda(\|\boldsymbol{\beta}_{h \mathcal{S}}\|_{1}-\|\hat{\boldsymbol{\theta}}_{h \mathcal{S}}+\boldsymbol{\beta}_{h \mathcal{S}}\|_{1}-\|\hat{\boldsymbol{\theta}}_{h \mathcal{S}^{C}}\|_{1}) \notag \\
			&\leq \lambda(\|\hat{\boldsymbol{\theta}}_{h \mathcal{S}}\|_{1}-\|\hat{\boldsymbol{\theta}}_{h \mathcal{S}^{C}}\|_{1}),
		\end{align} 
		where $\mathcal{S}=\{j \in [p]: {\beta}_{h, j} \neq 0\}$, and $\mathcal{S}^{C}=[p]/\mathcal{S}$.
		For $\text{I}_{2}=\nabla\mathcal{L}_{n}({\boldsymbol{\beta}}_{h})(\boldsymbol{\beta}_{h}-\hat{\boldsymbol{\beta}}_{h})$ in \eqref{the upper bound initial decomposition}, by H$\ddot{\text{o}}$lder's inequality and Lemma 6 in Supplementary Material,  we have 
		\begin{align}\label{I2 initial decomposition}
			\text{I}_{2}&=\nabla\mathcal{L}_{n}({\boldsymbol{\beta}}_{h})(\boldsymbol{\beta}_{h}-\hat{\boldsymbol{\beta}}_{h}) \notag \\
			&\leq \|\nabla\mathcal{L}_{n}({\boldsymbol{\beta}}_{h})\|_{\infty}\|\boldsymbol{\beta}_{h}-\hat{\boldsymbol{\beta}}_{h}\|_{1} \notag \\
			& \lesssim \sqrt{\frac{\log p}{n}}\|\hat{\boldsymbol{\theta}}_{h}\|_{1}.
		\end{align}
		Recall that  $\lambda\asymp \sqrt{(\log p)/n}$, and assume that $\lambda/2 >C\sqrt{(\log p)/n}$. By triangle and Cauchy-Schwartz inequalities, combining \eqref{I1 initial decomposition} with \eqref{I2 initial decomposition}, we have 
		\begin{align}\label{upper bound  final result}
			& \quad \langle\nabla\mathcal{L}_{n}(\hat{\boldsymbol{\beta}}_{h})- \nabla\mathcal{L}_{n}({\boldsymbol{\beta}}_{h}), \hat{\boldsymbol{\beta}}_{h}-\boldsymbol{\beta}_{h}\rangle \notag \\
			& \leq \lambda\left(\|\hat{\boldsymbol{\theta}}_{h \mathcal{S}}\|_{1}-\|\hat{\boldsymbol{\theta}}_{h \mathcal{S}^{C}}\|_{1}\right)+C\sqrt{\frac{\log p}{n}}\|\hat{\boldsymbol{\theta}}_{h}\|_{1} \notag \\
			& =\lambda(\|\hat{\boldsymbol{\theta}}_{h \mathcal{S}}\|_{1}-\|\hat{\boldsymbol{\theta}}_{h \mathcal{S}^{C}}\|_{1})+C\sqrt{\frac{\log p}{n}}(\|\hat{\boldsymbol{\theta}}_{h \mathcal{S}}\|_{1}+\|\hat{\boldsymbol{\theta}}_{h \mathcal{S}^{C}}\|_{1}) \notag \\
			&=\left\{\lambda+C\sqrt{\frac{\log p}{n}}\right\}\|\hat{\boldsymbol{\theta}}_{h \mathcal{S}}\|_{1}+\left\{C\sqrt{\frac{\log p}{n}}-\lambda\right\}\|\hat{\boldsymbol{\theta}}_{h \mathcal{S}^{C}}\|_{1} \notag \\
			%						&\leq \frac{3}{2}\lambda\|\hat{\boldsymbol{\theta}}_{h \mathcal{S}}\|_{1}-\frac{1}{2}\lambda\|\hat{\boldsymbol{\theta}}_{h \mathcal{S}^{C}}\|_{1} \notag \\
			%						&\leq \frac{3}{2}\lambda\|\hat{\boldsymbol{\theta}}_{h \mathcal{S}}\|_{1} \notag \\
			%						&\leq \frac{3}{2}\lambda\sqrt{s}\|\hat{\boldsymbol{\theta}}_{h \mathcal{S}}\|_{2}\notag \\
			& \leq \frac{3}{2}\lambda\sqrt{s}\|\hat{\boldsymbol{\beta}}_{h}-\boldsymbol{\beta}_{h}\|_{2}.
		\end{align}
		The inequality \eqref{upper bound  final result} shows that the right side of \eqref{aim inequality} holds. 
		
		Next, we focus on proving the left side of \eqref{aim inequality}.  Note that for any $\Delta \in \mathbb{R}^{p}$,
		\begin{align}
			\varepsilon_{n}(\Delta)=\left\{\nabla\mathcal{L}_{n}(\boldsymbol{\beta}_{h}+\Delta)-\nabla\mathcal{L}_{n}(\boldsymbol{\beta}_{h})\right\}^{\top}\Delta 
			=\frac{1}{n}(\hat{\boldsymbol{U}}\Delta)^{\top}(\hat{\boldsymbol{U}}\Delta). \notag
		\end{align}
		Because $\langle\nabla\mathcal{L}_{n}(\hat{\boldsymbol{\beta}}_{h})- \nabla\mathcal{L}_{n}({\boldsymbol{\beta}}_{h}), \hat{\boldsymbol{\beta}}_{h}-\boldsymbol{\beta}_{h}\rangle \geq 0$ in \eqref{upper bound  final result}, we have 
		$\|\hat{\boldsymbol{\theta}}_{\mathcal{S}^{C}}\|_{1} \leq 3\|\hat{\boldsymbol{\theta}}_{\mathcal{S}}\|_{1}$. Besides, we have the sparsity assumption that $s\{(\log p)/n+1/p\}\rightarrow 0$. Then, according to Lemma C.2 in \cite{fanJ2023}, we have 
		\begin{align}
			\Pr\left\{\frac{\left|\frac{1}{n}(\hat{\boldsymbol{U}}\hat{\boldsymbol{\theta}}_{h})^{\top}(\hat{\boldsymbol{U}}\hat{\boldsymbol{\theta}}_{h})\right|}{\|\hat{\boldsymbol{\theta}}_{h}\|_{2}^{2}}\geq C\right\}\rightarrow 1. \notag 
		\end{align}
		That is, with high probability, we have 
		\begin{align}\label{lower bound final result}
			\langle\nabla\mathcal{L}_{n}(\hat{\boldsymbol{\beta}}_{h})- \nabla\mathcal{L}_{n}({\boldsymbol{\beta}}_{h}), \hat{\boldsymbol{\beta}}_{h}-\boldsymbol{\beta}_{h}\rangle \geq C\|\hat{\boldsymbol{\beta}}_{h}-\boldsymbol{\beta}_{h}\|_{2}^{2},
		\end{align}
		which shows that the left side of \eqref{aim inequality} holds. 
		Therefore, in conclusion, we prove the $\ell_{2}$-norm error bound for $\hat{\boldsymbol{\beta}}_{h}$, \text{i.e.}
		\begin{align} \label{estimation l2 norm result}
			\|\hat{\boldsymbol{\beta}}_{h}-\boldsymbol{\beta}_{h}\|_{2}\leq C\lambda\sqrt{s},
		\end{align}
		where $\lambda\asymp \sqrt{(\log p)/n}$.
		Because $	\langle\nabla\mathcal{L}_{n}(\hat{\boldsymbol{\beta}}_{h})- \nabla\mathcal{L}_{n}({\boldsymbol{\beta}}_{h}), \hat{\boldsymbol{\beta}}_{h}-\boldsymbol{\beta}_{h}\rangle \geq 0$ in \eqref{upper bound  final result}, we have 
		\begin{align}
			\|\hat{\boldsymbol{\beta}}_{h}-\boldsymbol{\beta}_{h}\|_{1} \leq C\sqrt{s}\|\hat{\boldsymbol{\beta}}_{h}-\boldsymbol{\beta}_{h}\|_{2}. \notag 
		\end{align}
		Therefore, 
		\begin{align}
			\|\hat{\boldsymbol{\beta}}_{h}-\boldsymbol{\beta}_{h}\|_{1} \leq C\lambda s.
		\end{align}
		Here, $\lambda\asymp \sqrt{(\log p)/n}$.\\
		The proof is completed.
	\end{proof}
	\subsection{Proof of Theorem \ref{debias decomposition theorem} } 
	\begin{proof}
		Recall that  $\boldsymbol{m}=\{m_{1}(Y),\ldots,m_{p}(Y)\}^{\top}$, with  $m_{j}(y)=\mathbb{E}[(X_{1j}-\boldsymbol{f}_{1}^{\top}\boldsymbol{b}_{j})\{\mathbb{I}(Y\geq y)-F(Y)\}]$. Let $\boldsymbol{m}_i$ be the i.i.d copies of $\boldsymbol{m}$. Namely, $\boldsymbol{m}_i=\{m_{1}(Y_i),\ldots,m_{p}(Y_i)\}^{\top}$.
		%Denote $\boldsymbol{m}({\boldsymbol{Y}})=\{\sum_{i=1}^{n}m_{1}({Y_{i}}),\ldots,\sum_{i=1}^{n}m_{p}({Y_{i}})\}^{\top}$, with  $m_{j}(y)=\mathbb{E}[(X_{1j}-\boldsymbol{f}_{1}^{\top}\boldsymbol{b}_{j})\{\mathbb{I}(Y\geq y)-F(Y)\}]$.
		We have the following decomposition.
		\begin{align*}
			\sqrt{n}\left(\tilde{\boldsymbol{\beta}}_{h}-{\boldsymbol{\beta}}_{h}\right)=\boldsymbol{Z}_{b}+\boldsymbol{E}.\notag 
		\end{align*}
		Here, $\boldsymbol{Z}_{b}=n^{-1/2}\sum_{i=1}^n\boldsymbol{\Sigma}_{u}^{-1}\left({\boldsymbol{u}}_i e_{hi}+\boldsymbol{m}_i\right)$   and $\boldsymbol{E}=\sqrt{n}\left(\tilde{\boldsymbol{\beta}}_{h}-{\boldsymbol{\beta}}_{h}\right)-n^{-1/2}\sum_{i=1}^n\boldsymbol{\Sigma}_{u}^{-1}\left({\boldsymbol{u}}_ie_{hi}+\boldsymbol{m}_i\right)$. 
		By Lemma 9 in Supplementary Material, as long as $s=o(\sqrt{n}/\log p)$ and $\log p <<n^{{(1-q)}/{(3-q)}}$, we have  $\|\boldsymbol{E}\|_{\infty}=o_{\mathbb{P}}(1)$, where $q$ is given in Assumption \ref{assumption on the sparsity of the inverse of SIgmau}.
		The proof is completed.
	\end{proof}

	% Acknowledgements and Disclosure of Funding should go at the end, before appendices and references
	\iffalse
	\acks{All acknowledgements go at the end of the paper before appendices and references.
		Moreover, you are required to declare funding (financial activities supporting the
		submitted work) and competing interests (related financial activities outside the submitted work).
		More information about this disclosure can be found on the JMLR website.}
	\fi
	% Manual newpage inserted to improve layout of sample file - not
	% needed in general before appendices/bibliography.

\normalem
\bibliographystyle{apalike}
\bibliography{bibliography0718}

	\newpage
	\section*{Figures}
	\setcounter{figure}{0}
	\renewcommand{\thefigure}{\arabic{figure}}
	\begin{figure}[H]%
		\centering
		\includegraphics[width=0.9\textwidth]{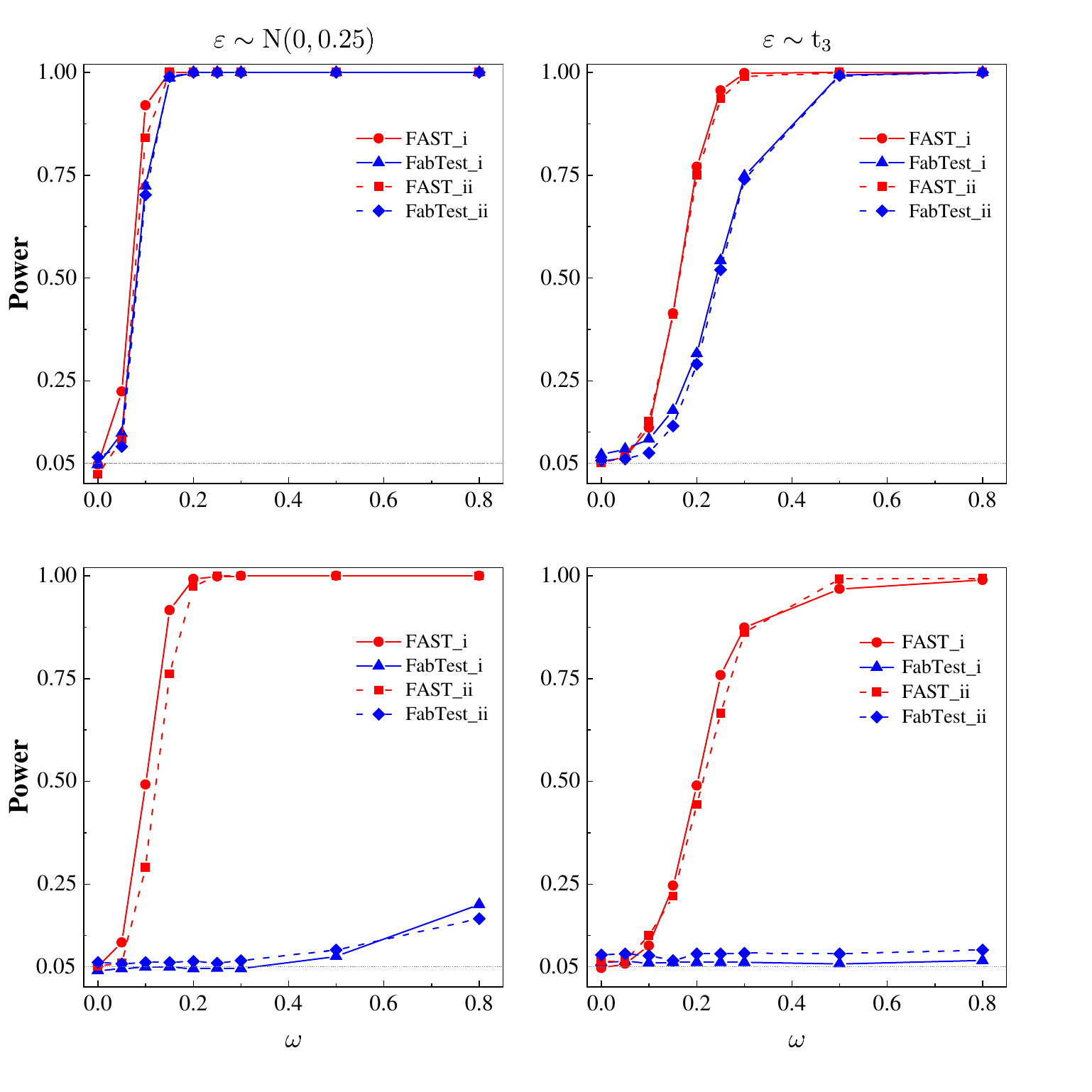}
		\caption{Power curves under  linear model \eqref{simulation generate data linear model} with $p=200$ and $\boldsymbol{\beta}=\omega\ast\left(\mathbf{1}_{3},\mathbf{0}_{p-3}\right)^{\top}$. 
			\iffalse
			The  ``FAST\_i'', ``FAST\_ii'', ``FabTest\_i'' and ``FabTest\_ii'' signify the results derived from the FASIM  and  FARM  corresponding to settings \eqref{generate F model1} and \eqref{generate F model2}  of $\boldsymbol{F}$ generation, respectively. 	The first row represent the results obtained with the original data, while the second row corresponds to the results of adding outliers.
			\fi
			The  ``FAST\_i'', ``FAST\_ii'', ``FabTest\_i'' and ``FabTest\_ii'' signify the results derived from the FAST in this paper   and  FabTest in \cite{fanJ2023}  corresponding to settings \eqref{generate F model1} and \eqref{generate F model2}  of $\boldsymbol{F}$ generation, respectively. 
			The first row represent the results obtained with the original data, while the second row corresponds to the results of adding outliers.
			The first column shows the results when the error follows {$\mathrm{N}(0, 0.25)$}, while the second column exhibits the results when the error follows  $\mathrm{t}_3$. }\label{Linear_P_200_our_Fan_figure}
	\end{figure}

	\begin{figure}[H]%
		\centering
		\includegraphics[width=0.9\textwidth]{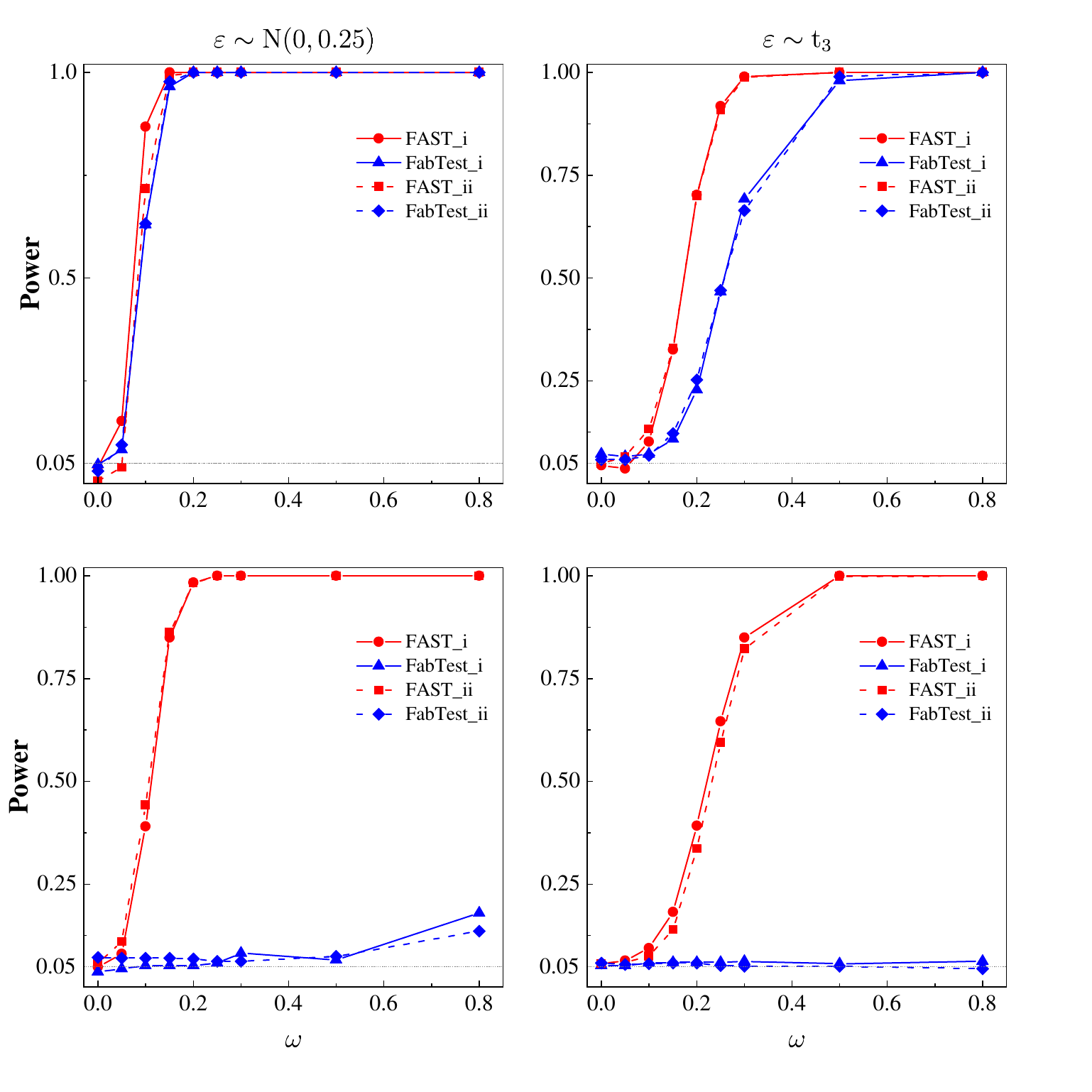}
		\caption{Power curves of linear model \eqref{simulation generate data linear model} with $p=500$ and $\boldsymbol{\beta}=\omega\ast\left(\mathbf{1}_{3},\mathbf{0}_{p-3}\right)^{\top}$.
			\iffalse
			The  ``FAST\_i'', ``FAST\_ii'', ``FabTest\_i'' and ``FabTest\_ii'' signify the results derived from the FASIM  and  FARM  corresponding to settings \eqref{generate F model1} and \eqref{generate F model2}  of $\boldsymbol{F}$ generation, respectively. 
			\fi
			The  ``FAST\_i'', ``FAST\_ii'', ``FabTest\_i'' and ``FabTest\_ii'' signify the results derived from the FAST in this paper   and  FabTest in \cite{fanJ2023}  corresponding to settings \eqref{generate F model1} and \eqref{generate F model2}  of $\boldsymbol{F}$ generation, respectively. 
			The first row represents the outcomes derived from the original data, while the second row corresponds to the results of adding outliers. The first column shows the outcomes assuming the error follows {$\mathrm{N}(0, 0.25)$}, while the second column exhibits the outcomes assuming the error follows  $\mathrm{t}_3$.}\label{Linear_P_500_our_Fan_figure}
	\end{figure}
	
	\begin{figure}[H]%
		\centering
		\includegraphics[width=0.9\textwidth]{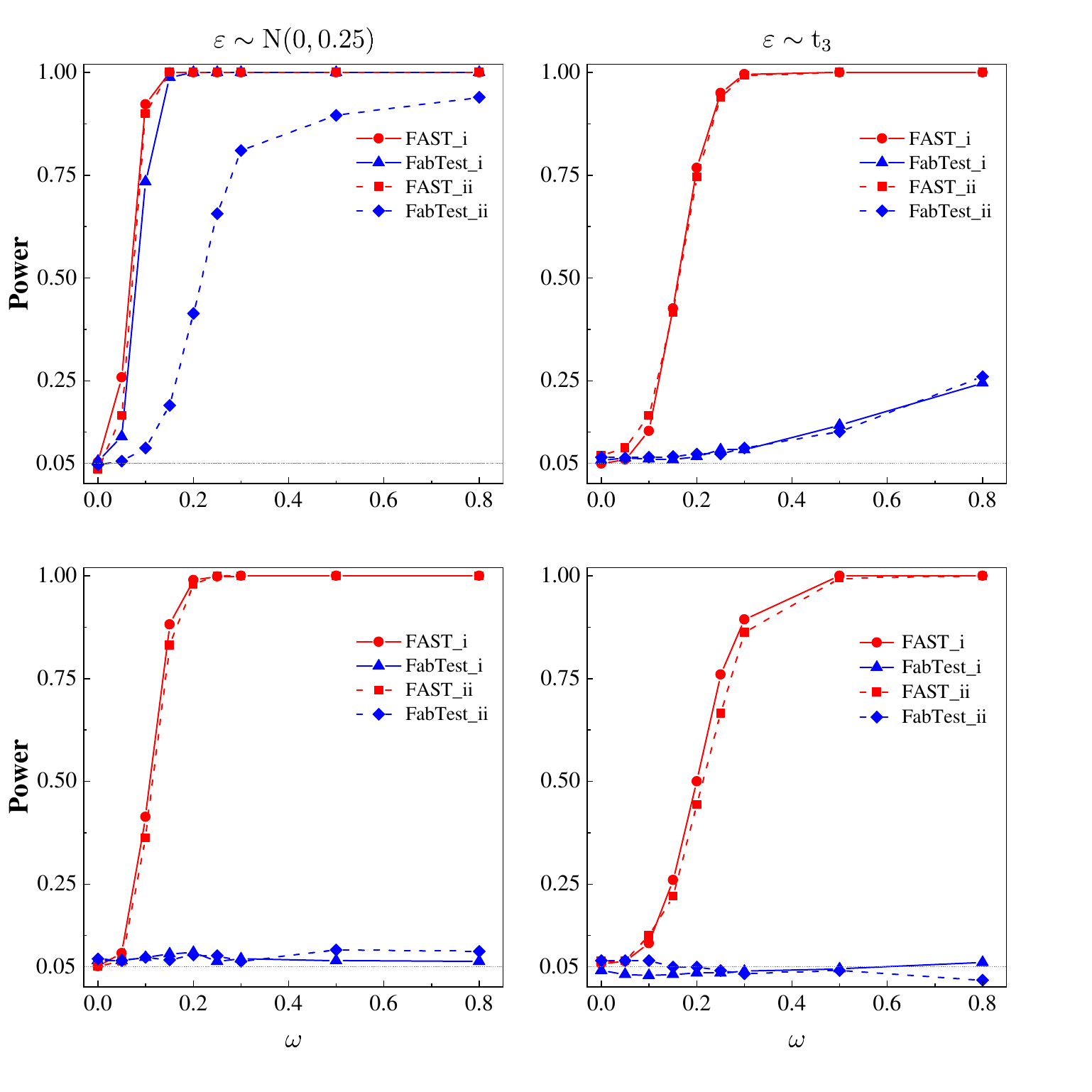}
		\caption{Power curves of non-linear model \eqref{simulation generate data Non-linear model} with $p=200$ and $\boldsymbol{\beta}=\omega\ast\left(\mathbf{1}_{3},\mathbf{0}_{p-3}\right)^{\top}$. 
			\iffalse
			The  ``FAST\_i'', ``FAST\_ii'', ``FabTest\_i'' and ``FabTest\_ii'' signify the results derived from the FASIM  and  FARM  corresponding to settings \eqref{generate F model1} and \eqref{generate F model2}  of $\boldsymbol{F}$ generation, respectively. 	\fi
			The  ``FAST\_i'', ``FAST\_ii'', ``FabTest\_i'' and ``FabTest\_ii'' signify the results derived from the FAST in this paper   and  FabTest in \cite{fanJ2023}  corresponding to settings \eqref{generate F model1} and \eqref{generate F model2}  of $\boldsymbol{F}$ generation, respectively. 
			The first row represents the outcomes derived from the original data, while the second row corresponds to the results of adding outliers. The first column shows the outcomes assuming the error follows {$\mathrm{N}(0, 0.25)$}, while the second column exhibits the outcomes assuming the error follows  $\mathrm{t}_3$.}\label{Non_Linear_P_200_our_Fan_figure}
	\end{figure}

	\begin{figure}[H]%
		\centering
		\includegraphics[width=0.9\textwidth]{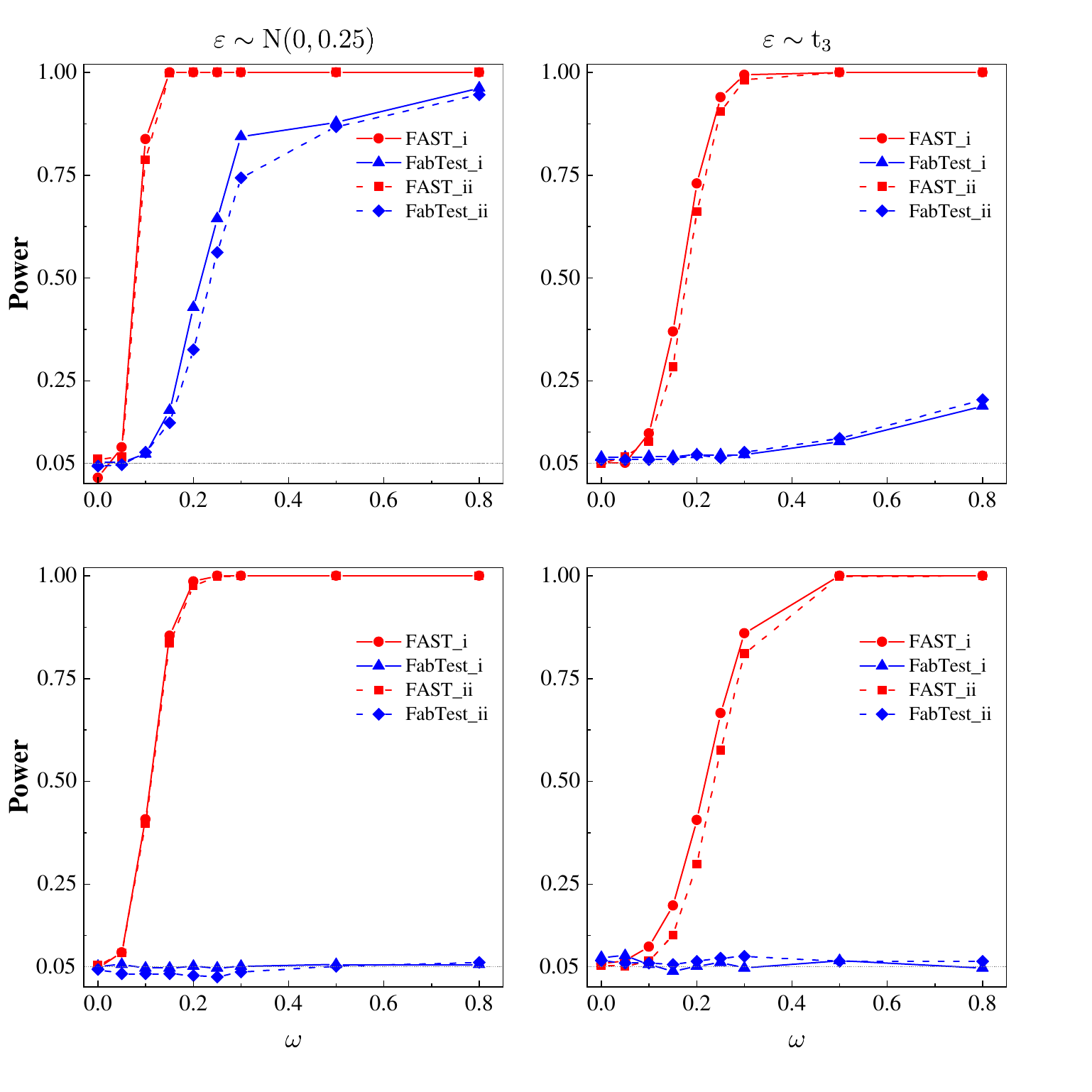}
		\caption{Power curves of non-linear model \eqref{simulation generate data Non-linear model} with $p=500$ and $\boldsymbol{\beta}=\omega\ast\left(\mathbf{1}_{3},\mathbf{0}_{p-3}\right)^{\top}$. 
			\iffalse
			The  ``FAST\_i'', ``FAST\_ii'', ``FabTest\_i'' and ``FabTest\_ii'' signify the results derived from the FASIM  and  FARM  corresponding to settings \eqref{generate F model1} and \eqref{generate F model2}  of $\boldsymbol{F}$ generation, respectively. 
			\fi
			The  ``FAST\_i'', ``FAST\_ii'', ``FabTest\_i'' and ``FabTest\_ii'' signify the results derived from the FAST in this paper   and  FabTest in \cite{fanJ2023}  corresponding to settings \eqref{generate F model1} and \eqref{generate F model2}  of $\boldsymbol{F}$ generation, respectively. 
			The first row represents the outcomes derived from the original data, while the second row corresponds to the results of adding outliers. The first column shows the outcomes assuming the error follows {$\mathrm{N}(0, 0.25)$}, while the second column exhibits the outcomes assuming the error follows  $\mathrm{t}_3$.}\label{Non_Linear_P_500_our_Fan_figure}
	\end{figure}
	
	\begin{figure}[H]
		\centering
		\includegraphics[width=0.9\textwidth]{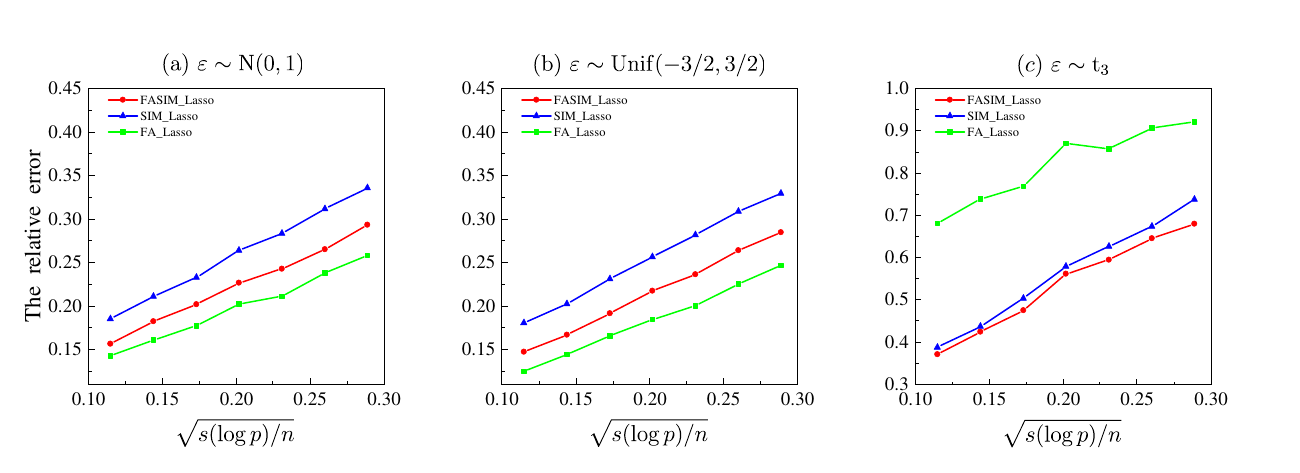}
		\caption{The relative errors of $\hat{\boldsymbol{\beta}}_{h}$ and $\hat{\boldsymbol{\beta}}$.
			\iffalse
			with  $L_{2}(\hat{\boldsymbol{\beta}}_{h}
			, \boldsymbol{\beta}_{h})=\|\hat{\boldsymbol{\beta}}_{h}-\boldsymbol{\beta}_{h}\|_{2}/\left\|\boldsymbol{\beta}_{h}\right\|_{2}$
			based on  500 replications\fi Figures (a), (b) and (c) depict the estimation results of linear model \eqref{simulation generate data linear model} with noise $\varepsilon$  following $\mathrm{N}(0, 1)$, $\mathrm{Unif}(-{3}/{2}, {3}/{2})$ and $\mathrm{t}_3$,   respectively. 
			{The ``FASIM\_Lasso'', ``SIM\_Lasso'' and ``FA\_Lasso''  represent the relative errors of parameter ${\boldsymbol{\beta}}_{h}$ under FASIM in this paper, SIM without incorporating the factor effect in \cite{rejchel2020rank}   and the  parameter ${\boldsymbol{\beta}}$ under FARM in \cite{fanJ2023}, respectively.}
			\iffalse
			In (a), (b) and (c), the red lines denote the estimation results using the method \eqref{empiricalestimator}  with data $(\hat{\boldsymbol{U}}, F_{n}(\boldsymbol{Y})-\frac{1}{2})$ (labeled as FA\_Lasso in the figures), and the blue lines represent the result using Lasso with data   $\left(\boldsymbol{X}, F_{n}(\boldsymbol{Y})-{1}/{2}\right)$ (labeled as Lasso in the figures).
			\fi}\label{l1_linear_Nonlinear_Gaussian_Uniform_figure}
	\end{figure}

	\begin{figure}[H]%
		\centering
		\includegraphics[width=0.9\textwidth]{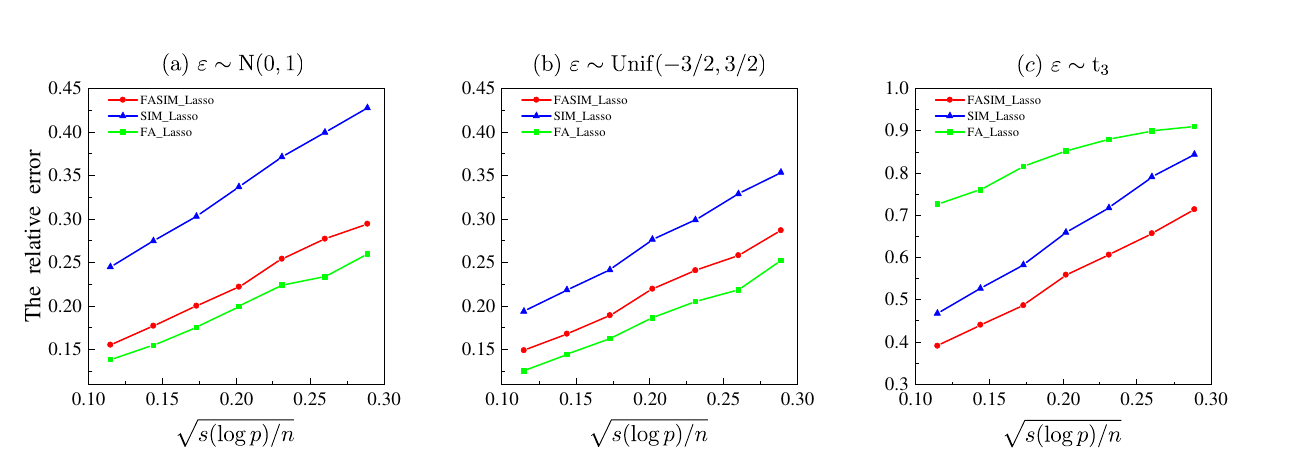}
		\caption{The relative errors of $\hat{\boldsymbol{\beta}}_{h}$ and $\hat{\boldsymbol{\beta}}$. 
			\iffalse
			with  
			$L_{2}(\hat{\boldsymbol{\beta}}_{h}, \boldsymbol{\beta}_{h})=\|\hat{\boldsymbol{\beta}}_{h}-\boldsymbol{\beta}_{h}\|_{2}/\left\|\boldsymbol{\beta}_{h}\right\|_{2}$ 
			based on 500 replications\fi Figures (a), (b) and (c) depict the estimation results of nonlinear model \eqref{simulation generate data Non-linear model} with noise $\varepsilon$  following $\mathrm{N}(0, 1)$, $\mathrm{Unif}(-{3}/{2}, {3}/{2})$ and $\mathrm{t}_3$,   respectively. {The ``FASIM\_Lasso'', ``SIM\_Lasso'' and ``FA\_Lasso''  represent the relative errors of parameter ${\boldsymbol{\beta}}_{h}$ under FASIM in this paper, SIM without incorporating the factor effect in \cite{rejchel2020rank}   and the  parameter ${\boldsymbol{\beta}}$ under FARM in \cite{fanJ2023}, respectively.}
			\iffalse
			In (a), (b) and (c), the red lines denote the estimation results using the method \eqref{empiricalestimator} with the data $(\hat{\boldsymbol{U}}, F_{n}(\boldsymbol{Y})-{1}/{2})$ (labeled as FA\_Lasso in the figures), and the blue lines represent the result using Lasso with data   $\left(\boldsymbol{X}, F_{n}(\boldsymbol{Y})-{1}/{2}\right)$ (labeled as Lasso in the figures).\fi
		}\label{l2_linear_Nonlinear_Gaussian_Uniform_figure}
		
	\end{figure}
	
	\begin{figure}[H]%
		\centering
		\includegraphics[width=0.9\textwidth]{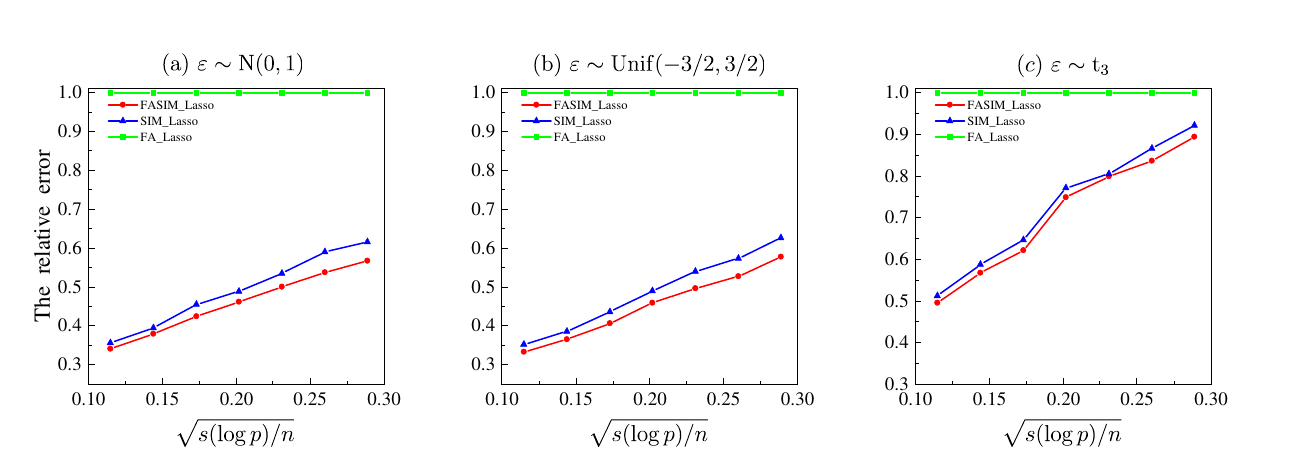}
		\caption{The relative errors for $\hat{\boldsymbol{\beta}}_{h}$ and $\hat{\boldsymbol{\beta}}$ 
			\iffalse
			with $L_{2}(\hat{\boldsymbol{\beta}}_{h}, \boldsymbol{\beta}_{h})=\|\hat{\boldsymbol{\beta}}_{h}-\boldsymbol{\beta}_{h}\|_{2}/\|\boldsymbol{\beta}_{h}\|_{2}$ and   $L_{2}(\hat{\boldsymbol{\beta}}, \boldsymbol{\beta})=\|\hat{\boldsymbol{\beta}}-\boldsymbol{\beta}\|_{2}/\|\boldsymbol{\beta}\|_{2}$ 
			\fi
			with outliers.  Figures (a), (b) and (c) depict the estimation results of linear model \eqref{simulation generate data linear model} with noise $\varepsilon$  following $\mathrm{N}(0, 1)$, $\mathrm{Unif}(-{3}/{2}, {3}/{2})$ and $\mathrm{t}_3$,   respectively.  {The ``FASIM\_Lasso'', ``SIM\_Lasso'' and ``FA\_Lasso''  represent the relative errors of parameter ${\boldsymbol{\beta}}_{h}$ under FASIM in this paper, SIM without incorporating the factor effect in \cite{rejchel2020rank}   and the  parameter ${\boldsymbol{\beta}}$ under FARM in \cite{fanJ2023}, respectively.}
			\iffalse
			In (a), (b) and (c), the red lines denote the estimation results using the method \eqref{empiricalestimator} with the data $(\hat{\boldsymbol{U}}, F_{n}(\boldsymbol{Y})-\frac{1}{2})$ in  this paper (labeled as FASIM in the figures), the blue lines denote the estimation results using the method with the data $(\hat{\boldsymbol{U}},\boldsymbol{Y}_{out})$ in \cite{fanJ2023} (labeled as FARM in the figures).\fi}\label{l1_ori_out__Linear_Nonlinear_Gaussian_Uniform_figure}
	\end{figure}

	\begin{figure}[H]%
		\centering
		\includegraphics[width=0.9\textwidth]{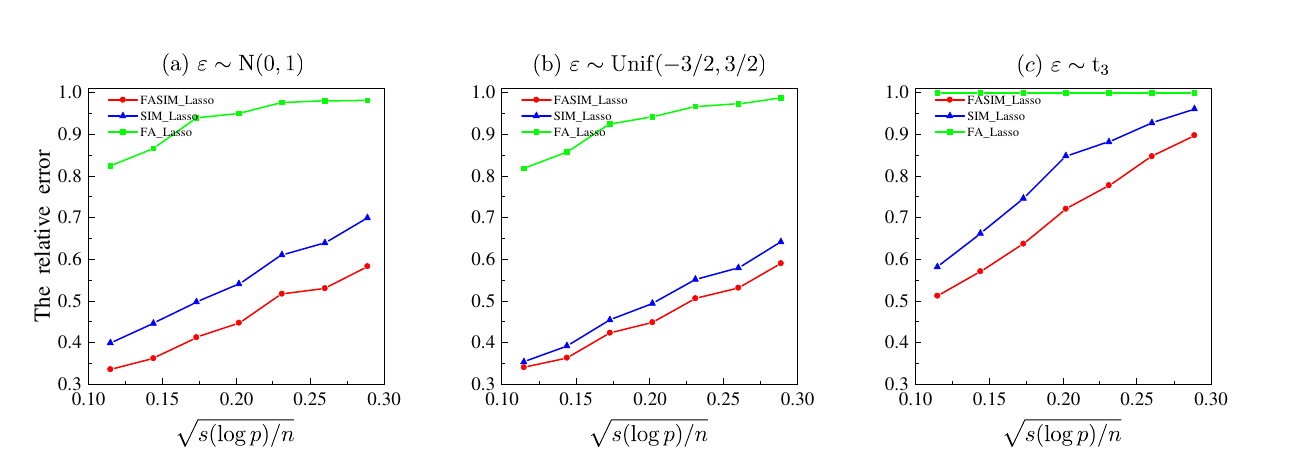}
		\caption{The relative errors of $\hat{\boldsymbol{\beta}}_{h}$ and $\hat{\boldsymbol{\beta}}$
			\iffalse
			with $L_{2}(\hat{\boldsymbol{\beta}}_{h}, \boldsymbol{\beta}_{h})=\|\hat{\boldsymbol{\beta}}_{h}-\boldsymbol{\beta}_{h}\|_{2}/\|\boldsymbol{\beta}_{h}\|_{2}$ and   $L_{2}(\hat{\boldsymbol{\beta}}, \boldsymbol{\beta})=\|\hat{\boldsymbol{\beta}}-\boldsymbol{\beta}\|_{2}/\|\boldsymbol{\beta}\|_{2}$ 
			\fi
			with outliers.  Figures (a), (b) and (c) depict the estimation results of nonlinear model \eqref{simulation generate data Non-linear model} with noise $\varepsilon$  following $\mathrm{N}(0, 1)$, $\mathrm{Unif}(-{3}/{2}, {3}/{2})$ and $\mathrm{t}_3$,  respectively.  {The ``FASIM\_Lasso'', ``SIM\_Lasso'' and ``FA\_Lasso''  represent the relative errors of parameter ${\boldsymbol{\beta}}_{h}$ under FASIM in this paper, SIM without incorporating the factor effect in \cite{rejchel2020rank}   and the  parameter ${\boldsymbol{\beta}}$ under FARM in \cite{fanJ2023}, respectively.}
			\iffalse
			In (a), (b) and (c), the red lines denote the estimation results using the method \eqref{empiricalestimator} with the data $(\hat{\boldsymbol{U}}, F_{n}(\boldsymbol{Y})-\frac{1}{2})$ in  this paper (labeled as FASIM in the figures), the blue lines denote the estimation results using the method with the data $(\hat{\boldsymbol{U}},\boldsymbol{Y}_{out})$  in \cite{fanJ2023} (labeled as FARM in the figures).
			\fi}\label{l2_ori_out__Linear_Nonlinear_Gaussian_Uniform_figure}
	\end{figure}

	\newpage
	\section*{Tables}
	\setcounter{table}{0} % 重置表格计数器
	\renewcommand{\thetable}{\arabic{table}} % 确保表格编号为阿拉伯数字格式
	\begin{table}[ht]
		\small
		
		\renewcommand\arraystretch{1.1}
		\centering \tabcolsep 12pt \LTcapwidth 6in
		\caption{The average computation time (Unit: second). The  ``FAST\_i'', ``FAST\_ii'', ``FabTest\_i'' and ``FabTest\_ii'' signify the results derived from the FAST   and  FabTest   corresponding to settings \eqref{generate F model1} and \eqref{generate F model2}  of $\boldsymbol{F}$ generation, respectively. 
			%The Setting\_i and Setting\_ii  represent scenarios where  the errors follow  $\mathrm{N}(0, 0.25)$ and $\mathrm{t}_3$ with the parameter dimension $p=200$, respectively. 
			%The Setting\_iii and Setting\_iv  represent scenarios where the parameter dimension $p=500$.
		}
		\label{computation time}
		\begin{threeparttable}
			\begin{tabular}{cccccc}
				\toprule
				$p$& $\varepsilon$&FAST\_i  & FabTest\_i  & FAST\_ii  & FabTest\_ii  \\ \midrule
				\multirow{2}{*}{$200$}&\multirow{1}{*}{$ \mathrm{N}(0, 0.25)$}       & 1.50      &  46.33     & 1.60    &  44.53    \\ 
				&\multirow{1}{*}{$\mathrm{t}_3$}           &  1.61      & 45.62     &  1.48    & 47.08\\ \hline
				
				\multirow{2}{*}{$500$}&\multirow{1}{*}{$\mathrm{N}(0, 0.25)$}          &  3.72     &  156.00    &  3.75   &  156.53   \\  
				&\multirow{1}{*}{$\mathrm{t}_3$}          &  3.65       & 208.86    &  3.73  &    146.21    \\ 
				\bottomrule
			\end{tabular}
		\end{threeparttable}
	\end{table}
	
	\begin{table}[H]
		\small
		
		\renewcommand\arraystretch{1.1}
		\centering \tabcolsep 12pt \LTcapwidth 6in
		\caption{The empirical coverage probabilities and average lengths of the confidence intervals.
		}
		\label{The coverage probability and average length of the confidence intervals.}
		\begin{threeparttable}
			\begin{tabular}{c|c|c|cccccc}
				\toprule
				\multicolumn{1}{c}{\multirow{1}{*}{Model}} &
				\multicolumn{1}{c}{\multirow{1}{*}{$p$}} & \multicolumn{1}{c}{\multirow{1}{*}{$\varepsilon$}} & $\text{CP}$  & $\text{CP}_{\mathcal{S}}$  &$\text{CP}_{\mathcal{S}^{C}}$   &  $\text{AL}$ & $\text{AL}_{\mathcal{S}}$  & $\text{AL}_{\mathcal{S}^{C}}$  \\ \midrule
				\multicolumn{1}{c}{\multirow{4}{*}{ Model 1}}&\multicolumn{1}{c}{\multirow{2}{*}{500}} & \multicolumn{1}{c}{$\mathrm{N}(0, 0.25)$}                  & 0.951  &  0.955 & 0.951  & 0.040 &0.039   & 0.040  \\
				\multicolumn{1}{l}{}&\multicolumn{1}{l}{}                  & \multicolumn{1}{c}{$ \mathrm{t}_3$}                  & 0.947  &0.950  &  0.947 &  0.066 & 0.064 & 0.066 \\ 
				\multicolumn{1}{l}{}&\multicolumn{1}{c}{\multirow{2}{*}{200}} & \multicolumn{1}{c}{$\mathrm{N}(0, 0.25)$}                  & 0.952  & 0.940  & 0.952  & 0.040  &0.039   &0.040  \\
				\multicolumn{1}{l}{}&\multicolumn{1}{l}{}                  & \multicolumn{1}{c}{$\mathrm{t}_3$}                  & 0.946  &0.943   & 0.946  &  0.067 &  0.066 & 0.067 \\ \hline
				\multicolumn{1}{c}{\multirow{4}{*}{ Model 2}}&\multicolumn{1}{c}{\multirow{2}{*}{500}} & \multicolumn{1}{c}{$ \mathrm{N}(0, 0.25)$}                  & 0.951  & 0.966 & 0.951  & 0.041  &  0.040 &  0.041\\
				\multicolumn{1}{c}{}   &\multicolumn{1}{c}{}                  & \multicolumn{1}{c}{$\mathrm{t}_3$}                  &0.946   & 0.930  &  0.946 &  0.069 & 0.068 & 0.069\\ 
				\multicolumn{1}{c}{}   &\multicolumn{1}{c}{\multirow{2}{*}{200}} & \multicolumn{1}{c}{$\mathrm{N}(0, 0.25)$}                  &0.951   &0.974   & 0.950  & 0.041  & 0.040  &0.041  \\
				\multicolumn{1}{c}{}   &\multicolumn{1}{l}{}                  & \multicolumn{1}{c}{$\mathrm{t}_3$}                 &0.947   & 0.956  & 0.947  &  0.067 & 0.066  & 0.067 \\ \bottomrule
			\end{tabular}
		\end{threeparttable}
	\end{table}

	\begin{table}[H]
		\small
		
		\renewcommand\arraystretch{1.1}
		\centering \tabcolsep 12pt \LTcapwidth 6in
		\caption{{The $p$-values for the original dataset and polluted dataset   in two different time periods. }}
		\label{$p$-values of each test for the real data based on FASIM  and the FARM.}
		\begin{threeparttable}
			\begin{tabular}{cccc}
				\toprule
				Time Period & Data    & F\_SIM   & F\_LM     \\ \midrule
				\multirow{2}{*}{1992.02-2007.10} &original     & 0.0000 & $2 \times 10^{-3}$\\
				&polluted & 0.0000 &0.0000 \\ \hline
				\multirow{2}{*}{2010.08-2020.02} &original    & 0.0000 & $1.7 \times 10^{-2}$ \\
				&polluted & 0.0000 & 0.2015 \\ 
				\bottomrule
			\end{tabular}
		\end{threeparttable}
	\end{table}
	
	\begin{table}[H]
		\small
		
		\renewcommand\arraystretch{1.1}
		\centering \tabcolsep 12pt \LTcapwidth 6in
		\caption{{The MSE  for the original dataset and polluted dataset  in different time periods.}}
		\label{Accuracy of  FASIM, FARM and SIM based on real data.}
		\begin{threeparttable}
			\begin{tabular}{cccc}
				\toprule
				Time Period                      & Data                     & FARM  & FASIM   \\ \midrule
				\multirow{2}{*}{1992.02-2007.10}  & \multirow{1}{*}{original}       &  0.2620       &  0.2764        \\ 
				& \multirow{1}{*}{polluted}           &  1.0592        &   0.9251     \\ \hline
				
				\multirow{2}{*}{2010.08-2020.02}& \multirow{1}{*}{original}          &  0.3328      &  0.3748     \\  
				& \multirow{1}{*}{polluted}          &  0.9432       &  0.8764             \\ 
				\bottomrule
			\end{tabular}
		\end{threeparttable}
	\end{table}

\end{document}